\documentclass[journal]{IEEEtran}
\IEEEoverridecommandlockouts

\usepackage{cite}
\usepackage{algorithm}
\usepackage[noend]{algpseudocode}
\usepackage{amsmath,amssymb,amsfonts}
\usepackage[pdftex]{graphicx}
\usepackage{subcaption}
\usepackage{textcomp}
\usepackage{xcolor}
\usepackage{pifont}
\usepackage{multirow}
\usepackage{makecell}
\usepackage{array}
\usepackage{diagbox}
\usepackage{soul}

\begin{document}

\title{Semantic MIMO Systems for Speech-to-Text Transmission}
\author{Zhenzi Weng,~\IEEEmembership{Member,~IEEE,} Zhijin Qin,~\IEEEmembership{Senior Member,~IEEE}, Huiqiang Xie,~\IEEEmembership{Member,~IEEE,}\\ Xiaoming Tao,~\IEEEmembership{Senior Member,~IEEE,} and Khaled B. Letaief,~\IEEEmembership{Fellow,~IEEE}
\thanks{Part of the work was presented at IEEE ICC Workshop 2023~\cite{10283658}.}
\thanks{This work was supported in part by the National Natural Science Foundation of China (NSFC) under Grant 62293484, 61925105, 62227801, and 62401227. This work was supported in part by the Nantong Research Institute for Advanced Communication Technologies and in part by the Fundamental Research Funds for the Central Universities under Grant 21624349. This work was also supported in part by the Hong Kong Research Grants Council under the Areas of Excellence scheme grant AoE/E-601/22-R. (Corresponding authors: Zhijin Qin and Xiaoming Tao.)}
\thanks{Zhenzi Weng is with Imperial College London, London SW7 2AZ, U.K. (e-mail: z.weng@imperial.ac.uk).}
\thanks{Zhijin Qin and Xiaoming Tao are with Tsinghua University, Beijing 100084, China. They are also with the State Key Laboratory of Space Network and Communications and the Beijing National Research Center for Information Science and Technology, Beijing, China (e-mail: qinzhijin@tsinghua.edu.cn; taoxm@tsinghua.edu.cn).}
\thanks{Huiqiang Xie is with Jinan University, Guangzhou 510632, China (e-mail: huiqiangxie@jnu.edu.cn).}
\thanks{Khaled B. Letaief is with The Hong Kong University of Science and Technology, Hong Kong, and also with the Peng Cheng Laboratory, Shenzhen 518066, China (e-mail: eekhaled@ust.hk).}
}

\maketitle
\begin{abstract}
Semantic communications have been utilized to execute numerous intelligent tasks by transmitting task-related semantic information instead of bits. In this article, we propose a semantic-aware speech-to-text transmission system for the single-user multiple-input multiple-output (MIMO) and multi-user MIMO communication scenarios, named SAC-ST. Particularly, a semantic communication system to serve the speech-to-text task at the receiver is first designed, which compresses the semantic information and generates the low-dimensional semantic features by leveraging the transformer module. In addition, a novel semantic-aware network is proposed to facilitate transmission with high semantic fidelity by identifying the critical semantic information and guaranteeing its accurate recovery. Furthermore, we extend the SAC-ST with a neural network-enabled channel estimation network to mitigate the dependence on accurate channel state information and validate the feasibility of SAC-ST in practical communication environments. Simulation results will show that the proposed SAC-ST outperforms the communication framework without the semantic-aware network for speech-to-text transmission over the MIMO channels in terms of the speech-to-text metrics, especially in the low signal-to-noise regime. Moreover, the SAC-ST with the developed channel estimation network is comparable to the SAC-ST with perfect channel state information.

\end{abstract}


\section{Introduction}
Semantic communications are the second level of communications defined by Shannon and Weaver~\cite{weaver1953recent} to deliver the semantic exchange and optimize the communication systems at the semantic level. To alleviate the scarcity of bandwidth resources and the explosion of data traffic, the semantic communication problem has been listed as one of the promising techniques for future intelligent communications~\cite{9770094,10183789,qin2021semantic,10639525}.

The conventional communications, i.e., the first level of communications, convert the input message into a bit sequence and aim to achieve a low bit-error rate (BER) by leveraging proper coding and decoding algorithms. The bit-oriented transmission framework has undergone a thriving development, but the attention to semantic information has been neglected. The ultimate goal of semantic communications is to exploit the semantic information inside the source and facilitate the semantic transmission to convey the meaning of the transmitter to the receiver. However, due to the limitation of a unified theory to represent the semantics of different sources and a mathematical formula to quantify the semantic information, the investigation of semantic communications has experienced a few decades of stagnation. Thanks to the success of artificial intelligence (AI) in recent years, the deep learning (DL)-enabled semantic communication architecture has been proposed and has shown its great potential to tackle the technical challenges across many aspects of conventional communications, e.g., resource allocation~\cite{sun2018learning} and physical layer communications~\cite{10183794}. Such success is attributed to the learning and fitting capabilities of diverse neural networks, which break the constraints of a mathematical model to exchange semantic information.

The DL-enabled semantic communication paradigm requires the least semantic ambiguity between the source and the recovered messages. It maps the source message into the low-dimensional semantic features, which reduces the volume of transmission data and mitigates the network traffic without performance degradation. Inspired by this, a preliminary research work of DL-enabled semantic communications for text transmission has been developed in~\cite{9398576}. This work designed a transformer-powered joint semantic-channel codecs mechanism, named DeepSC, to minimize the semantic error by measuring the meaning difference between the input and recovered sentences. Inspired by DeepSC, Peng~\emph{et al.}~\cite{10486856} further proposed R-DeepSC to improve the semantic fidelity and against the semantic impairments inherent in the sentence through a semantic error corrector. In~\cite{9791409}, a text semantic communication system is developed to ensure transmission reliability by adopting the hybrid automatic repeat request (HARQ) scheme. Liang~\emph{et al.}~\cite{9814575} studied a reasoning-based semantic communication paradigm to represent semantics through a graph-based knowledge architecture and a semantic interpretation framework. Xie~\emph{et al.} investigated task-oriented semantic communications for text transmission in~\cite{9830752} and introduced a memory component in~\cite{10159023} to tackle the transmission problems in dynamic channel environments. Nam~\emph{et al.}~\cite{10287475} developed a novel sequential semantic communication system for text-to-image generation through multi-modal generation techniques. Moreover, a DL-enabled semantic communication system for speech signals, named DeepSC-S, has been designed in~\cite{9450827} by extracting and transmitting global semantic information. Wei~\textit{et al.}~\cite{10001043} devised a real-time audio semantic communication system to strengthen the audio quality and alleviate the physical channel attenuation by capturing the long-distance dependency. Han~\textit{et al.}~\cite{9953316} proposed an efficient speech semantic communication system by introducing the redundancy removal module to drop insignificant semantics and reduce the transmission data. In~\cite{10094680}, a deep speech semantic transmission system is developed to further improve the intelligibility of the reconstructed speech by leading to a flexible rate-distortion trade-off. Weng~\emph{et al.}~\cite{10038754} proposed a task-oriented semantic communication system for speech transmission, named DeepSC-ST, which dramatically outperforms the conventional speech transmission paradigm. In~\cite{10431795}, a unified semantic communication system for multimodal data has been proposed by fusing shared trainable parameters amongst various tasks to control the dimension of extracted semantics and reduce the transmission data.

At the same time, semantic communications for image/video transmission have attracted numerous research interests\cite{9953076,9796572,10101778,10107714,9953099,9953071,cicchetti2024language,10531769,10599525,9955991,9953110,10105154,10198383}. Particularly, a generative adversarial network (GAN)-enabled semantic communication system for image transmission has been proposed in~\cite{9953076} by interpreting the image meaning. Xu~\emph{et al.}~\cite{9796572} developed a joint image transmission and scene classification scheme in a reinforcement learning-enabled semantic communication system. A robust image semantic transmission scheme has been studied in~\cite{10101778} by incorporating a masked vector quantized-variational autoencoder to combat semantic error and improve image classification accuracy. Nan~\emph{et al.}~\cite{10107714} designed a training efficiency mechanism for image semantic communication by employing a distributed learning architecture to balance the trade-off between computation capability and user privacy. More recently, Cicchetti~\emph{et al.}~\cite{cicchetti2024language} established a semantic communication framework that combines text with compressed image embeddings and leverages a diffusion model to reconstruct images. Zhao~\emph{et al.}~\cite{10531769} developed a large language model-powered semantic communication system to improve the reconstructed image quality by transmitting multimodal features and Qiao~\emph{et al.}~\cite{10599525} introduced a latency-aware semantic communication mechanism to enable low-rate and low-latency visual transmission with high semantic fidelity. A video semantic transmission network is established in~\cite{9955991}, which utilizes a semantic detector at the receiver to reduce the semantic error. Thanks to the advancements in computer vision, the semantic communication for performing video tasks, extended reality (XR), has been designed in~\cite{10198383}, in which the insignificant semantic information is highly compressed to mitigate the transmission traffic.

DL-enabled semantic communications have also been investigated in multiple-input multiple-output (MIMO) transmission scenarios\cite{10597355,10304536,yao2022versatile,9921202,9943000,10510413}. Wu~\emph{et al.}~\cite{10597355} proposed a vision transformer-enabled semantic communication system for image transmission over MIMO channels by using a self-attention module to exploit the contextual semantic features and the channel conditions. In~\cite{10304536}, an orthogonal space-time block code (OSTBC) is utilized to improve the image transmission robustness to MIMO channel variants, and an equalizer is leveraged to estimate the codewords. Yao~\emph{et al.}~\cite{yao2022versatile} designed a versatile semantic coded transmission scheme over MIMO fading channels to adaptively adjust the coding rate allocation and stream mapping. Additionally, Luo~\emph{et al.}~\cite{9921202} considered a semantic precoding-aided MIMO system to alleviate the effect of physical channels by a new multimodal information fusion scheme. A massive MIMO semantic transmission system for image restoration is developed in~\cite{9943000}, in which the post-filter and image statistics are leveraged to improve the quality of reconstructed images. In~\cite{10510413}, Zhang~\emph{et al.} established a channel feedback-enabled semantic communication framework in MIMO systems to ensure the image reconstruction and proposed a metric, termed semantic distortion outage probability, to measure the reliability of semantic transmission.

In this article, we devise a semantic-aware speech-to-text transmission scheme in single-user MIMO (SU-MIMO) and multi-user MIMO (MU-MIMO) communication systems, named SAC-ST. We argue that previous state-of-the-art (SOTA) works on semantic communications for speech transmission, e.g., SemAudio~\cite{10001043} and DeepSC-ST~\cite{10038754}, have limited contributions to the analysis of semantic information, and the attention mechanisms for extracting the semantic features are not fully exploited. To address these challenges, a transformer-enabled semantic communication system for speech-to-text transmission over MIMO channels, named DeepSC-ST 2.0, is first developed. The DeepSC-ST 2.0 is evolved from the DeepSC-ST. Particularly, in DeepSC-ST 2.0, we redesign the semantic coding mechanism to avoid the intermediate spectrum and replace recurrent neural network (RNN) modules in DeepSC-ST with transformer modules, further eliminating the semantic redundancy and exploring the long-dependence of semantic information to improve the system performance compared to the DeepSC-ST, especially when contending with lengthy speech sequences. Moreover, based on the singular value decomposition (SVD) precoding of MIMO channels, we investigate the inequality of semantic information and utilize the SNR difference of decomposed equivalent SISO subchannels to boost the high semantic fidelity transmission by transmitting critical semantics over SISO subchannels with high SNRs. A novel semantic-aware network tailored for such a semantic transmission mechanism is proposed to identify the importance of different semantics and return the corresponding coefficients. The contributions of this paper are summarized as follows:
\begin{itemize}
\item A semantic communication system for speech-to-text transmission over MIMO fading channels, named DeepSC-ST 2.0, is developed to serve users only requesting text information by leveraging the convolutional neural network (CNN) to condense the input speech sequence and the transformer to extract the low-dimensional semantic features. The proposed scheme avoids the additional spectrum extraction and achieves superior speech-to-text transmission performance compared to the existing literature~\cite{10038754,9953316}.

\item To improve the semantic fidelity, we propose a novel neural network-enabled semantic-aware network, which is incorporated with DeepSC-ST 2.0 to form the proposed SAC-ST. According to the inequality of semantic features and the inconsistent SNRs of multiple independent SISO subchannels based on the SVD precoding of MIMO channels, the semantic-aware network perceives the essential semantics and grants them high priority to be recovered accurately over good subchannels.

\item To validate the feasibility of SAC-ST in practical communication scenarios, a novel channel precoding algorithm is carried out to reduce the computational complexity in the MU-MIMO environment, and a lightweight MIMO channel estimation network is jointly designed with the SAC-ST to minimize the dependence on perfect channel state information (CSI) in the SU-MIMO and MU-MIMO transmission paradigms.

\item Simulation results verify the superiority of the SAC-ST compared to the framework without the semantic-aware network. Moreover, the SAC-ST with the ChanEst network achieves remarkable performance similar to the SAC-ST with the perfect channel estimation.
\end{itemize}

The rest of this article is organized as follows. In Section~\uppercase\expandafter{\romannumeral2}, the motivations of the proposed system are introduced. Section~\uppercase\expandafter{\romannumeral3} presents the details of the SAC-ST and the ChanEst network. Simulation results are discussed in Section~\uppercase\expandafter{\romannumeral4} and Section~\uppercase\expandafter{\romannumeral5} draws our conclusions.

\emph{Notation}: The single boldface letters represent vectors or matrices. $\boldsymbol y\in\mathfrak R^{M\times N}$ and $\boldsymbol y\in\mathfrak C^{M\times N}$ are two matrices with real values and complex values, respectively, and their size is $M\times N$. $\boldsymbol y^T$ and $\boldsymbol y^H$ and indicates the transpose and conjugate transpose of matrix $\boldsymbol y$, respectively. ${\mathbf D}_{M\times N}$ denotes the $M\times N$ matrix with non-zero diagonal elements and all other elements equal to zero. ${\mathbf O}_{M\times N}$ is the $M\times N$ zero matrix. $\mathrm{mod}(M,\;N)$ returns the modulo operation of $M$ by $N$.

\section{Motivations}
This section briefly describes the MIMO transmission system with SVD precoding. We will then discuss the motivations of the proposed semantic-aware communication system over SU-MIMO channels and MU-MIMO channels.

\subsection{MIMO Transmission with SVD Precoding}
\begin{figure}[tbp]
\centering
\includegraphics[width=1.0\linewidth]{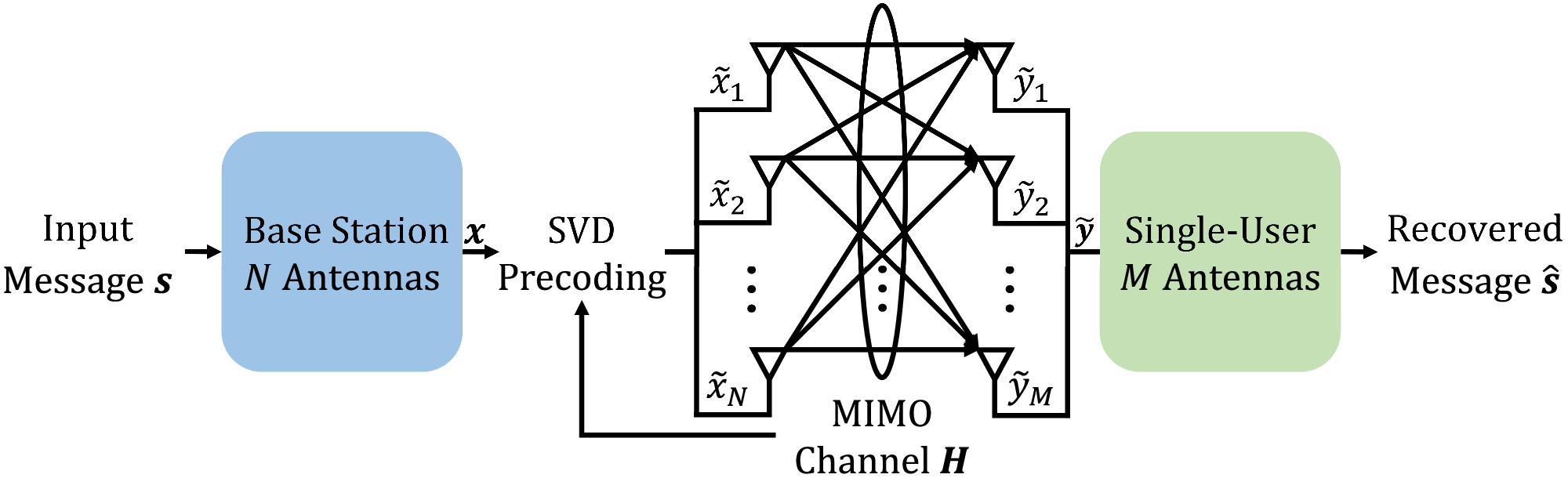} 
\caption{Conventional SU-MIMO communication system with SVD precoding.}
\label{SU-MIMO SVD precoding}
\end{figure}

The conventional SU-MIMO transmission architecture with SVD precoding is shown in Fig.~\ref{SU-MIMO SVD precoding}, where the base station (BS) and the single-user have $N$ antennas and $M$ antennas, respectively. From the figure, the input message, $\boldsymbol s$, is encoded to the symbols, $\boldsymbol x=\left[x_1,x_2,\dots,x_N\right]^T$. According to the SVD of MIMO channel, symbols $\boldsymbol x$ are converted into the precoded symbols, $\widetilde{\boldsymbol x}=\left[{\widetilde x}_1,{\widetilde x}_2,\dots,{\widetilde x}_N\right]^T$, that are distributed to $N$ antennas for transmission. The user with $M$ antennas receives the symbols, $\boldsymbol y=\left[y_1,y_2,\dots,y_M\right]^T$, and produces the recovered message, $\widehat{\boldsymbol s}$.

Denote the MIMO channel as $\boldsymbol H\in\mathfrak C^{M\times N}$, the SVD operation of $\boldsymbol H$ can then be expressed as follows,
\begin{equation}
\boldsymbol H=\boldsymbol U\boldsymbol\Lambda\boldsymbol V^H,
\label{SU-MIMO channel SVD}
\end{equation}
where $\boldsymbol U\in\mathfrak C^{M\times M}$ and $\boldsymbol V^H\in\mathfrak C^{N\times N}$ are the unitary matrices, $\boldsymbol\Lambda\in{\mathbf D}_{M\times N}$ contains the non-negative singular values as diagonal elements, i.e., $\lbrack\Lambda_{11},\Lambda_{22},\dots,\Lambda_{QQ}\rbrack$, where $Q=\min(N,M)$. Note that the singular values are listed in descending order, i.e., $\Lambda_{11}>\Lambda_{22}>\cdots>\Lambda_{QQ}$.

Based on the SVD of $\boldsymbol H$, the precoding process from $\boldsymbol x$ to $\widetilde{\boldsymbol x}$ can be expressed as $\widetilde{\boldsymbol x}=\boldsymbol V\boldsymbol x$, then the received symbols, $\boldsymbol y$ can be obtained from $\widetilde{\boldsymbol x}$ after passing through the MIMO channel, denoted as
\begin{equation}
\begin{split}
\boldsymbol y&=\boldsymbol H\widetilde{\boldsymbol x}+\boldsymbol n \\
&=\boldsymbol U\boldsymbol\Lambda\boldsymbol V^H\boldsymbol V\boldsymbol x+\boldsymbol n \\
&=\boldsymbol U\boldsymbol\Lambda\boldsymbol x+\boldsymbol n,
\label{SU-MIMO channel 1}
\end{split}
\end{equation}
where $\boldsymbol n$ indicates independent and identically distributed (i.i.d.) Gaussian noise with a SNR value, $\sigma$. By introducing $\boldsymbol U^H$, (\ref{SU-MIMO channel 1}) can be expressed as
\begin{equation}
\boldsymbol U^H\boldsymbol y=\boldsymbol\Lambda\boldsymbol x+\boldsymbol U^H\boldsymbol n.
\label{SU-MIMO channel 2}
\end{equation}

Let $\widetilde{\boldsymbol y}=\boldsymbol U^H\boldsymbol y$ and $\widetilde{\boldsymbol n}=\boldsymbol U^H\boldsymbol n$, then (\ref{SU-MIMO channel 2}) can be rewritten as
\begin{equation}
\widetilde{\boldsymbol y}=\boldsymbol\Lambda\boldsymbol x+\widetilde{\boldsymbol n}.
\label{SU-MIMO channel 3}
\end{equation}

According to the diagonal property of the matrix $\boldsymbol\Lambda$, (\ref{SU-MIMO channel 3}) is equivalent to
\begin{equation}
\begin{bmatrix}{\widetilde y}_1\\{\widetilde y}_2\\\vdots\\{\widetilde y}_Q\end{bmatrix}=\begin{bmatrix}\Lambda_{11}x_1+{\widetilde n}_1\\\Lambda_{22}x_2+{\widetilde n}_2\\\vdots\\\Lambda_{QQ}x_Q+{\widetilde n}_Q\end{bmatrix}.
\label{SU-MIMO channel 4}
\end{equation}

The SVD precoding-based MIMO channel can be treated as multiple independent SISO subchannels, and the noise SNRs of these SISO subchannels are inconsistent due to the difference of singular values, denoted as
\begin{equation}
\begin{split}
\widetilde{\boldsymbol\sigma}&=\lbrack{\widetilde\sigma}_1,{\widetilde\sigma}_2,\dots,{\widetilde\sigma}_q\rbrack \\
&=\lbrack\left(\Lambda_{11}\right)^2,\left(\Lambda_{22}\right)^2,\dots,\left(\Lambda_{QQ}\right)^2\rbrack\boldsymbol\ast\sigma,
\label{new snr}
\end{split}
\end{equation}
where the first SISO subchannel has the highest SNR value of $\left(\Lambda_{11}\right)^2\ast\sigma$ and exhibits minimal noise impairment.

\subsection{Motivations of MIMO Semantic Transmission}
\begin{figure}[tbp]
\centering
\begin{minipage}[t]{0.8\linewidth}
\centering
\includegraphics[width=1.0\textwidth]{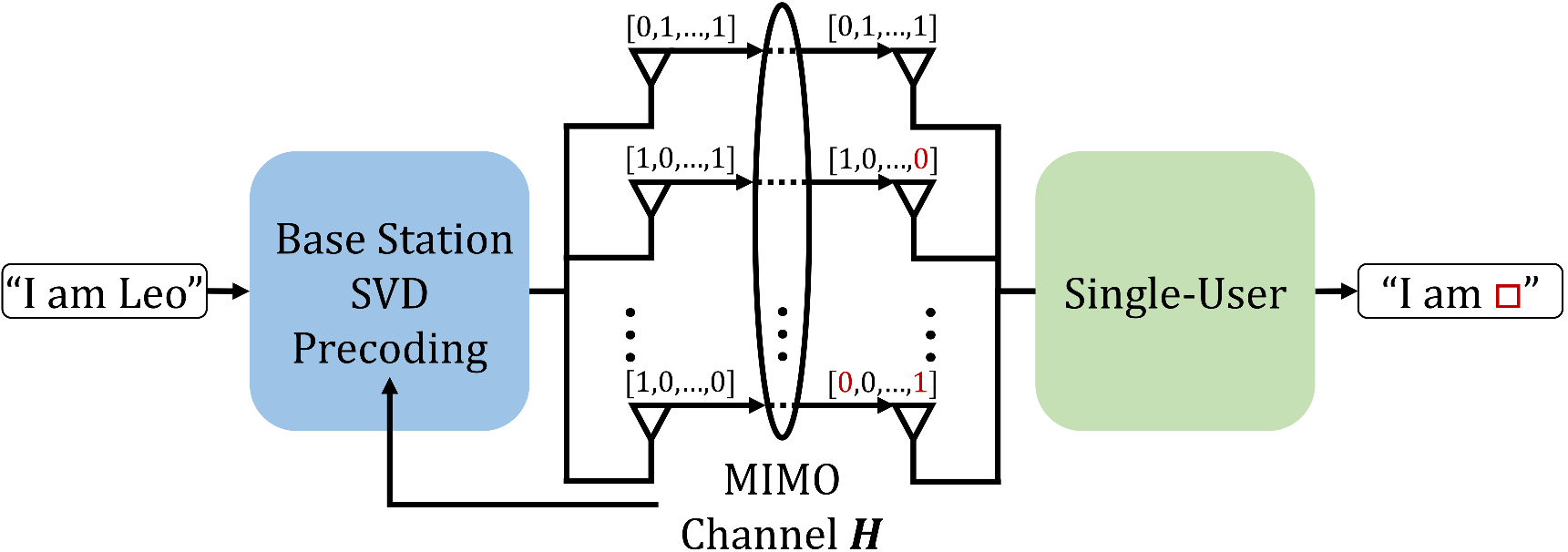}
\subcaption{Conventional communication system}
\label{MIMO bits transmission}
\end{minipage}
\begin{minipage}[t]{0.8\linewidth}
\centering
\includegraphics[width=1.0\textwidth]{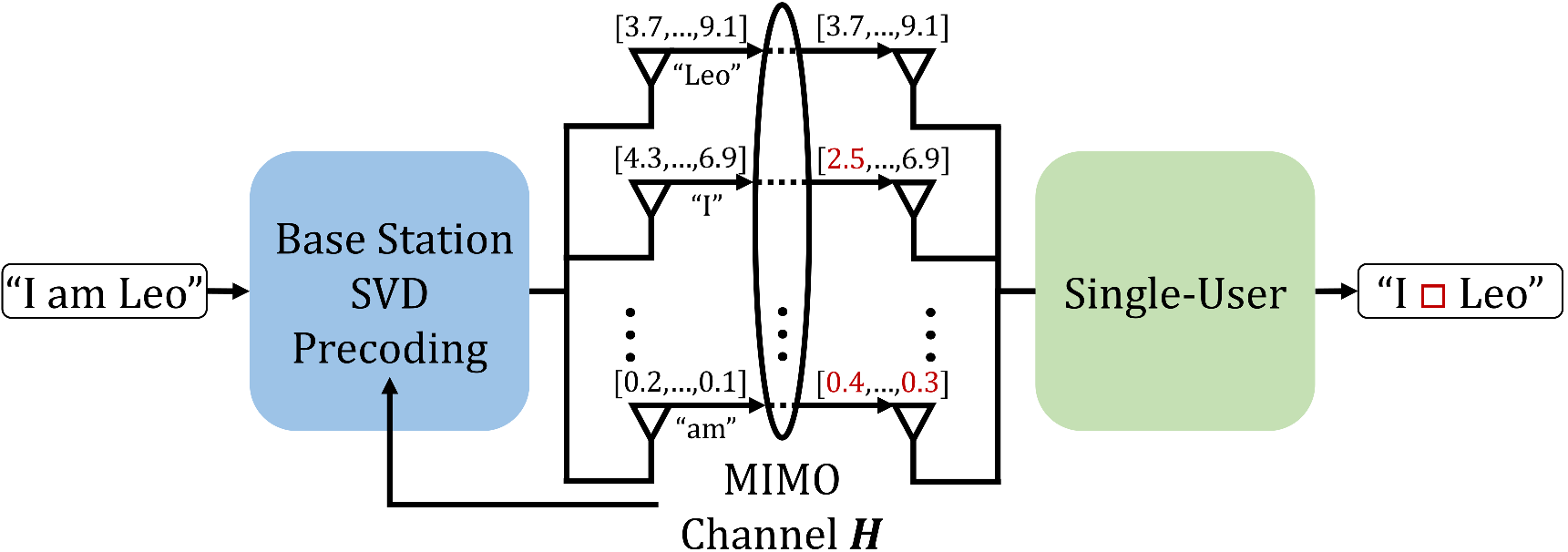}
\subcaption{Semanitc communication system}
\label{MIMO semantics transmission}
\end{minipage}
\caption{An example of the conventional and semantic SU-MIMO communication systems with SVD precoding.}
\label{MIMO bits and semantics transmission}
\end{figure}

In conventional communication systems, the input message is mapped into the bit sequence, as shown in Fig.~\ref{MIMO bits and semantics transmission} (a). From the figure, the bits transmitted over different SISO subchannels are considered equally important and contain no semantics. In this context, the key information ``Leo'' may be corrupted in the recovered message. The semantic communication paradigm converts the input message into semantic features and the information contained by different semantic features is inconsistent, indicating the importance of various features is unequally distributed. Inspired by this, the semantic features with high importance are identified and placed into the SISO subchannels with high SNRs prior to transmission. By doing so, the essential semantic information is received accurately to maximize the semantic fidelity of the recovered message. As shown in Fig.~\ref{MIMO bits and semantics transmission} (b), the semantic features corresponding to the key information ``Leo'' are placed into the top SISO subchannel with the highest SNR, while the semantic features corresponding to the less critical information ``am'' are transmitted over the SISO subchannel with the lowest SNR, which preserves the key information ``Leo'' and ensures the intelligibility of the obtained sentence. However, an efficient algorithm to distinguish the importance of the extracted semantic features and a sort operation to these semantic features are necessary to perform such a semantic transmission mechanism.

The semantic importance-based transmission approach is also applicable to the MU-MIMO communication scenarios. Assume that the number of receiver users is $K$, with each user having $M$ antennas, and the number of BS antennas is $N$, where $N$ is a multiple of $K$, $KM\geq N>M$. The encoded symbols are $\boldsymbol x=\left[\boldsymbol x_1^T,\boldsymbol x_2^T,\dots,\boldsymbol x_K^T\right]^T$, where ${\boldsymbol x}_k=\left[x_{k,1},x_{k,2}\dots,x_{k,N/K}\right]^T$, $k=\left[1,\dots,K\right]$, is the information requested by the $k$-th user. The MIMO channel of $K$ users is $\boldsymbol H\in\mathfrak C^{\left(K\ast M\right)\times N}$. Denote the channel corresponding to the $k$-th user as ${\boldsymbol H}_k=\left[{\boldsymbol H}_{k,1},{\boldsymbol H}_{k,2}\dots,{\boldsymbol H}_{k,K}\right]$, where ${\boldsymbol H}_k\in\mathfrak C^{M\times N}$ and ${\boldsymbol H}_{k,k}\in\mathfrak C^{M\times\left(N/K\right)}$, then the symbols received by the $k$-th user can be denoted as
\begin{equation}
\begin{split}
{\boldsymbol y}_k&={\boldsymbol H}_k\boldsymbol x+{\boldsymbol n}_k \\
&={\boldsymbol H}_{k,k}{\boldsymbol x}_k+\sum_{k^\ast}{\boldsymbol H}_{k,k^\ast}{\boldsymbol x}_{k^\ast}+{\boldsymbol n}_k,
\label{MU-MIMO channel 1}
\end{split}
\end{equation}
where $k^\ast=\left[1,\dots,k-1,k+1,\dots K\right]$. $\displaystyle\sum_{k^\ast}{\boldsymbol H}_{k,k^\ast}{\boldsymbol x}_{k^\ast}$ is the multi-user interference (MUI) and ${\boldsymbol n}_k$ is the Gaussian noise for the $k$-th user.

To achieve the transmission mechanism that prioritizes semantic importance, the precoding operation performed to symbols $\boldsymbol x$ is responsible for eliminating the MUI and enabling the MIMO channel decomposition to attain equivalent multiple SISO subchannels. The conventional block diagonalization (BP) algorithm~\cite{1259395} is proficient in accomplishing this objective. However, it requires each receiver user to possess knowledge of the CSI pertaining to all other users and demands a considerable computational burden due to the necessity of two SVD precoding operations, including the first one to obtain a precoding matrix, $\boldsymbol V^{\mathrm{B}}$, that completely removes the MUI and the second one to decompose the channel of each user into parallel subchannels according to a precoding matrix $\boldsymbol V^{\mathrm{P}}$. Then the joint precoding matrix of the BP algorithm is $\boldsymbol V^{\mathrm{BP}}=\boldsymbol V^{\mathrm{B}}\boldsymbol V^{\mathrm{P}}$. Toward this end, we propose a novel precoding algorithm to overcome the limitations of the BP algorithm. Particularly, the MIMO channel ${\boldsymbol H}_k$ is decomposed according to the SVD operation as follows,
\begin{equation}
\begin{split}
{\boldsymbol H}_k&={\boldsymbol U}_k{\boldsymbol\Lambda}_k\left({\boldsymbol V}_k\right)^H \\
&={\boldsymbol U}_k{\boldsymbol\Lambda}_k\left[\boldsymbol V_{k,1}^{\sim\mathrm{zero}},\boldsymbol V_{k,2}^{\mathrm{zero}},\boldsymbol V_{k,3}^{\mathrm{zero}},\dots,\boldsymbol V_{k,K}^{\mathrm{zero}}\right]^H,
\label{MU-MIMO channel SVD}
\end{split}
\end{equation}
where ${\boldsymbol\Lambda}_k\in{\mathbf D}_{M\times N}$ contains $M$ non-negative singular values as diagonal elements, i.e., $\lbrack\Lambda_{k,11},\Lambda_{k,22},\dots,\Lambda_{k,MM}\rbrack$. $\boldsymbol V_{k,1}^{\sim\mathrm{zero}}\in\mathfrak C^{N\times\left(N/K\right)}$ is in the non-null space of matrix ${\boldsymbol H}_k$ and $\boldsymbol V_{k,k^{'}}^{\mathrm{zero}}\in\mathfrak C^{N\times\left(N/K\right)}$, $k^{'}=\left[2,\dots,K\right]$, belongs to the null space of ${\boldsymbol H}_k$. In other words, $\left({\boldsymbol V}_k\right)^H\boldsymbol V_{k,1}^{\sim\mathrm{zero}}={\mathbf D}_{N\times\left(N/K\right)}$ and its diagonal elements are all equal to one, and ${\boldsymbol H}_k\boldsymbol V_{k,k^{'}}^{\mathrm{zero}}={\mathbf O}_{M\times\left(N/K\right)}$. Then, the precoded symbols $\widetilde{\boldsymbol x}$ is expressed as
\begin{equation}
\widetilde{\boldsymbol x}=\boldsymbol V_{k,1}^{\sim\mathrm{zero}}{\boldsymbol x}_k+\sum_{k^{'},k^\ast}\boldsymbol V_{k,k^{'}}^{\mathrm{zero}}{\boldsymbol x}_{k^\ast}.
\label{MU-MIMO SVD precoding}
\end{equation}

The received symbols of the $k$-th user, ${\boldsymbol y}_k$, is then obtained from the precoded symbols $\widetilde{\boldsymbol x}$ according to the precoding operation, denoted as
\begin{equation}
\begin{split}
{\boldsymbol y}_k&={\boldsymbol H}_k\widetilde{\boldsymbol x}+{\boldsymbol n}_k \\
&={\boldsymbol H}_k\left(\boldsymbol V_{k,1}^{\sim\mathrm{zero}}{\boldsymbol x}_k+\sum_{k^{'},k^\ast}\boldsymbol V_{k,k^{'}}^{\mathrm{zero}}{\boldsymbol x}_{k^\ast}\right)+{\boldsymbol n}_k \\
&={\boldsymbol U}_k{\boldsymbol\Lambda}_k\left({\boldsymbol V}_k\right)^H\boldsymbol V_{k,1}^{\sim\mathrm{zero}}{\boldsymbol x}_k+{\boldsymbol n}_k \\
&={\boldsymbol U}_k{\boldsymbol\Lambda}_k{\boldsymbol x}_k+{\boldsymbol n}_k.
\label{MU-MIMO channel 2}
\end{split}
\end{equation}

Let ${\widetilde{\boldsymbol y}}_k=\left({\boldsymbol U}_k\right)^H{\boldsymbol y}_k$ and ${\widetilde{\boldsymbol n}}_k=\left({\boldsymbol U}_k\right)^H{\boldsymbol n}_k$, then (\ref{MU-MIMO channel 2}) can be written as
\begin{equation}
{\widetilde{\boldsymbol y}}_k={\boldsymbol\Lambda}_k{\boldsymbol x}_k+{\widetilde{\boldsymbol n}}_k.
\label{MU-MIMO channel 3}
\end{equation}

By doing so, the MUI is completely removed and the MIMO channel, ${\boldsymbol H}_k$, behaves as multiple independent SISO subchannels to deliver the information of the $k$-th user, which inspires to assign the essential information to the good SISO subchannels to achieve the high semantic fidelity transmission. However, when transmitting the symbols of the $k$-th user through (\ref{MU-MIMO channel 3}), the precoded symbols of other users still undergo the channel fading and the MUI. To ensure overall semantic fidelity for $K$ users, we transmit the insignificant information of other users when the objective is to recover the information of the $k$-th user accurately.

\section{Semantic-Aware Speech-to-Text Transmission Over MIMO Systems with Channel Estimation}
This section introduces the details of the proposed SAC-ST. We consider the SU-MIMO and MU-MIMO transmission schemes. Besides, a channel estimation network is developed to eliminate the reliance on accurate CSI and deployed in both the SU-MIMO and MU-MIMO communication scenarios.

\subsection{SU-MIMO Communicaton System}
\begin{figure}[tbp]
\centering
\begin{minipage}[t]{0.45\linewidth}
\centering
\includegraphics[width=1.0\textwidth]{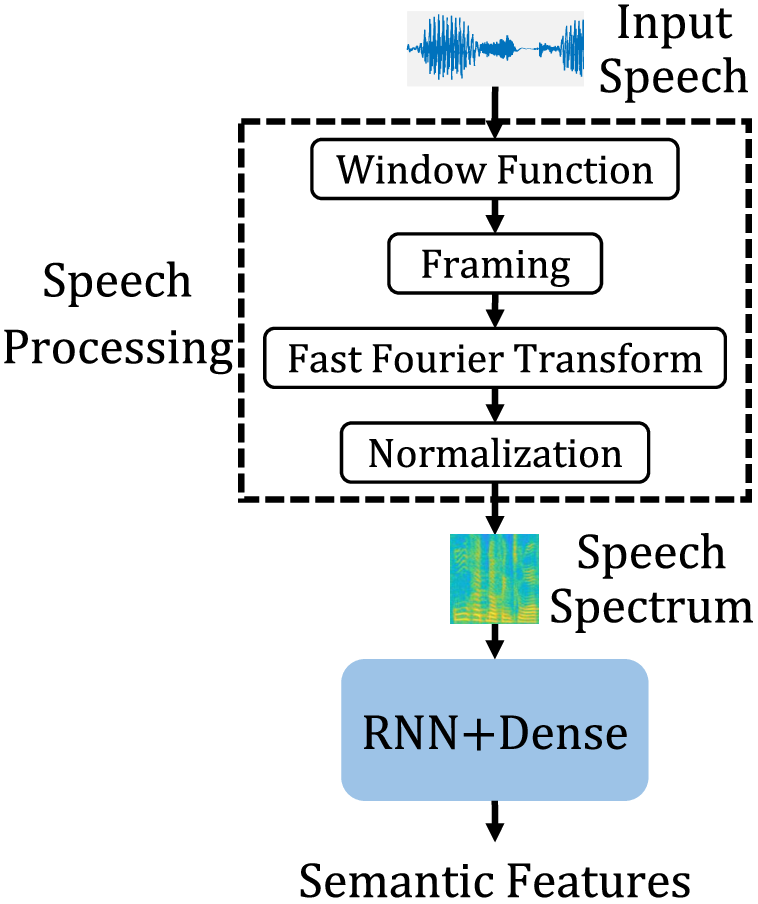}
\subcaption{}
\label{semantic encoder 1}
\end{minipage}
\begin{minipage}[t]{0.45\linewidth}
\centering
\includegraphics[width=1.0\textwidth]{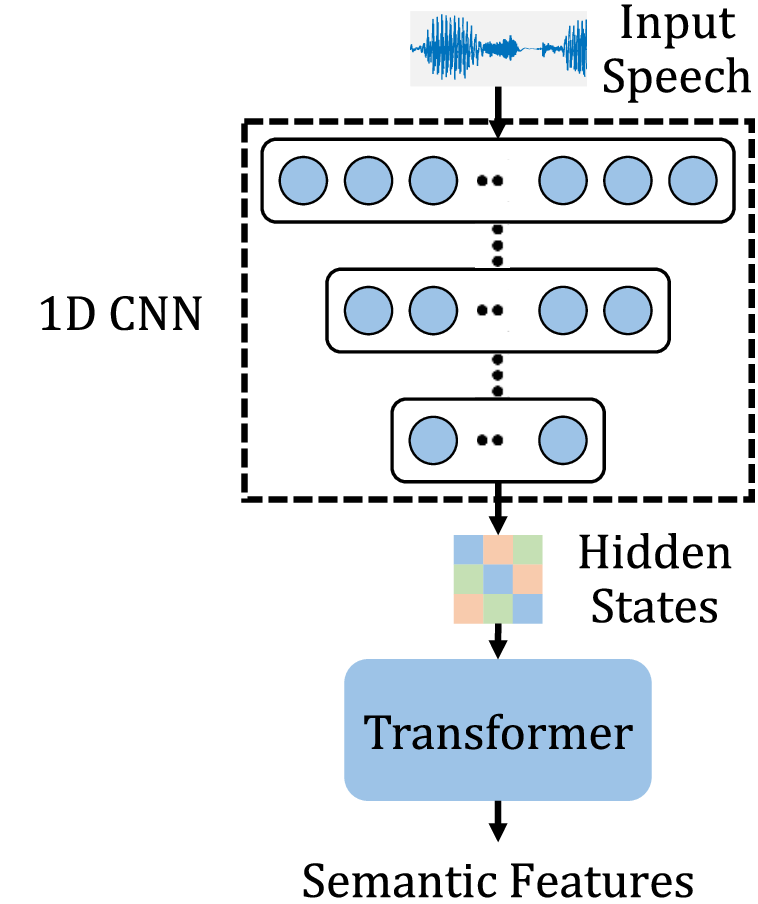}
\subcaption{}
\label{semantic encoder 2}
\end{minipage}
\caption{The model structure of (a) the semantic encoder in the existing works (b) and the proposed semantic encoder.}
\label{proposed semantic encoder}
\end{figure}

During the SU-MIMO transmission, the data transmitted over multiple SISO subchannels is aggregated to one user at the receiver. In the proposed SAC-ST, we design the DeepSC-ST 2.0 to serve the speech-to-text task. It is noteworthy that the existing works~\cite{10038754,9953316} require a pre-processing step to input speech signals before neural networks, which runs counter to intelligent semantic extraction and hinders the end-to-end gradient backpropagation to the original speech signals. To address this, the one-dimensional (1D) CNN is adopted to directly condense speech signals into intermediate representations, and the transformer is leveraged to explore the semantics inherent in these representations, as shown in Fig.~\ref{proposed semantic encoder}. Additionally, a semantic-aware network is developed to quantify the importance of the semantic features extracted by the semantic encoder, as shown in Fig.~\ref{proposed SAC-ST for SU-MIMO}. From the figure, the first and second training stages are performed to train the DeepSC-ST 2.0 and the semantic-aware network, respectively, which are adopted in the testing stage to achieve the semantic transmission with high semantic fidelity by identifying the essential semantic features and transmitting them over the SISO subchannels with high SNRs.

\subsubsection{\textbf{First training stage}}
\begin{figure*}[tbp]
\centering
\begin{minipage}[t]{1.0\linewidth}
\centering
\includegraphics[width=1.0\textwidth]{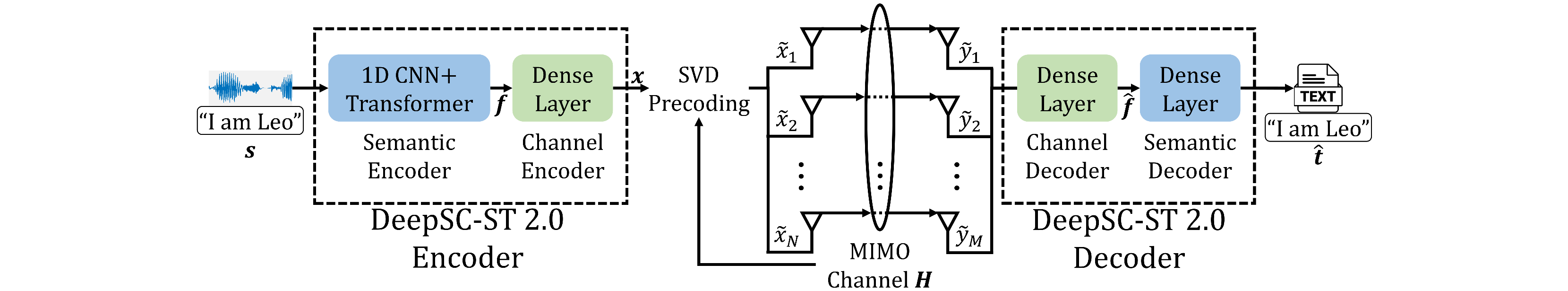}
\subcaption{First training stage: Train DeepSC-ST 2.0.}
\label{first training stage}
\vspace{0.2cm}  
\end{minipage}
\begin{minipage}[t]{1.0\linewidth}
\centering
\includegraphics[width=1.0\textwidth]{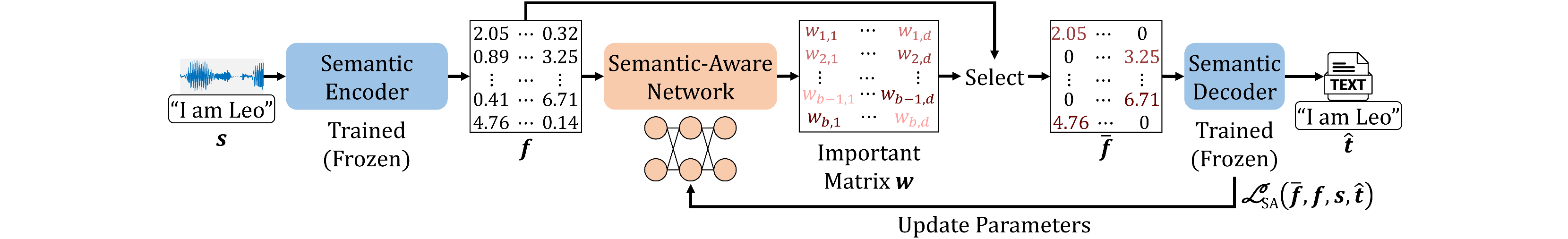}
\subcaption{Second training stage: Train semantic-aware network.}
\label{second training stage}
\vspace{0.2cm}  
\end{minipage}
\begin{minipage}[t]{1.0\linewidth}
\centering
\includegraphics[width=1.0\textwidth]{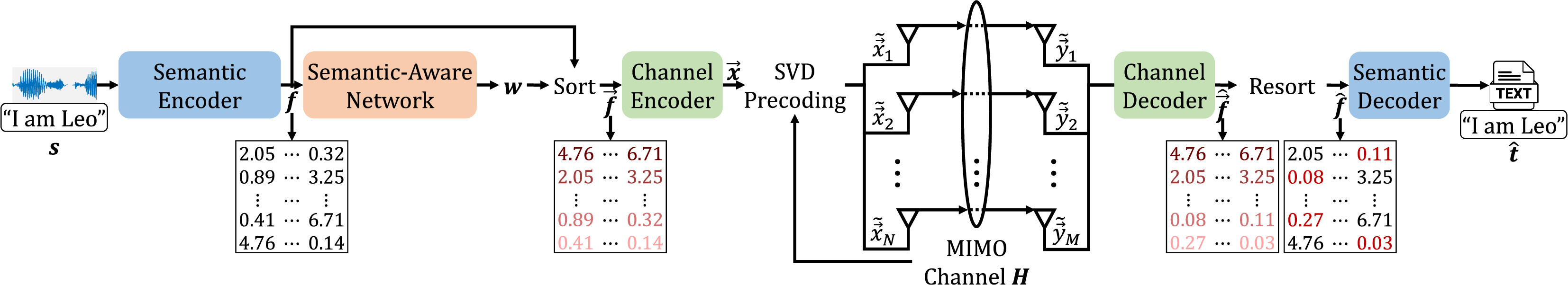}
\subcaption{Testing stage: Test with DeepSC-ST 2.0 and semantic-aware network.}
\label{testing stage}
\end{minipage} 
\caption{The model structure of the proposed semantic-aware speech-to-text transmission system (SAC-ST) for SU-MIMO transmission with SVD precoding.}
\label{proposed SAC-ST for SU-MIMO}
\end{figure*}

The goal of the first training stage is to train the DeepSC-ST 2.0 that recovers the text sequence at the receiver from the input speech over MIMO channels with SVD precoding. As shown in Fig.~\ref{proposed SAC-ST for SU-MIMO} (a), the 1D CNN and transformer-enabled semantic encoder converts the input speech, $\boldsymbol s$, into the low-dimensional semantic features, $\boldsymbol f$. The channel encoder processes $\boldsymbol f$ and produces the symbols, $\boldsymbol x$, that are precoded into $\widetilde{\boldsymbol x}$ to be transmitted over the MIMO channel, $\boldsymbol H$. Therefore, $\widetilde{\boldsymbol x}$ can be denoted as
\begin{equation}
\widetilde{\boldsymbol x}=\boldsymbol V{\mathfrak T}_{\mathrm C}({\mathfrak T}_{\mathrm S}(\boldsymbol s))\;\;\;\;\mathrm w.\mathrm r.\mathrm t.\;\;\;\boldsymbol\alpha,
\label{transmitter}
\end{equation}
where ${\mathfrak T}_{\mathrm S}(\cdot)$ and ${\mathfrak T}_{\mathrm C}(\cdot)$ indicate the semantic encoder and the channel encoder, respectively. $\boldsymbol\alpha$ is their trainable parameters. $\boldsymbol V$ is the precoding matrix obtained through the SVD of $\boldsymbol H$.

$\widetilde{\boldsymbol x}$ is transmitted to the user over the MIMO channel according to (\ref{SU-MIMO channel 1}). As aforementioned that the SVD precoding-based MIMO channel can be regarded as multiple SISO subchannels. The received symbols, $\widetilde{\boldsymbol y}$, are passing through the channel decoder to obtain the recovered semantic features, $\widehat{\boldsymbol f}$. Finally, the semantic decoder takes $\widehat{\boldsymbol f}$ as the input and produces the text sequence, $\widehat{\boldsymbol t}$, to serve the users only requesting the text information. The process to obtain $\widehat{\boldsymbol t}$ from $\widetilde{\boldsymbol y}$ can be expressed as
\begin{equation}
\widehat{\boldsymbol t}=\mathfrak T_{\mathrm S}^{-1}(\mathfrak T_{\mathrm C}^{-1}(\widetilde{\boldsymbol y}))\;\;\;\;\mathrm w.\mathrm r.\mathrm t.\;\;\;\boldsymbol\beta,
\label{receiver}
\end{equation}
where $\mathfrak T_{\mathrm C}^{-1}(\cdot)$ and $\mathfrak T_{\mathrm S}^{-1}(\cdot)$ represent the channel decoder and the semantic decoder, respectively. $\boldsymbol\beta$ is their trainable parameters.

The connectionist temporal classification (CTC)~\cite{graves2006connectionist} is adopted as the loss function to update all the trainable parameters of DeepSC-ST 2.0, ${\boldsymbol\theta}_1=(\boldsymbol\alpha,\boldsymbol\beta)$, denoted as
\begin{equation}
{\mathcal L}_{\mathrm{CTC}}({\boldsymbol\theta}_1)=-\ln\sum_{A\in\mathfrak A(\boldsymbol s,\boldsymbol\;\boldsymbol t)}p\left(\left.\widehat{\boldsymbol t}=\boldsymbol t\right|\boldsymbol s\right),
\label{CTC loss SU-MIMO}
\end{equation}
where $\mathfrak A(\boldsymbol s,\boldsymbol\;\boldsymbol t)$ is all the alignments from $\boldsymbol t$ to $\boldsymbol s$ and $\boldsymbol t$ is the correct text sequence. Assume $\boldsymbol t=\left[\mathrm t,\;\mathrm h,\;\mathrm i,\;\mathrm s\right]$, the valid alignments involve $\left[blank,\;\mathrm t,\;blank,\;\mathrm h,\;\mathrm i,\;\mathrm s,\;blank\right]$, or $\left[\mathrm t,\;blank,\;\mathrm h,\;\mathrm i,\;blank,\;blank,\;\mathrm s\right]$, etc, because the $blank$ token is eliminated when attaining $\widehat{\boldsymbol t}$.

The training algorithm of the DeepSC-ST 2.0 is described in Algorithm \ref{first training algorithm SU-MIMO}.

\subsubsection{\textbf{Second training stage}}
A semantic-aware network is developed in the second training stage to analyse the semantic features and return their importance. Particularly, the trained semantic encoder from the first training stage is leveraged to process input speech $\boldsymbol s$ and produce semantic features $\boldsymbol f$. Then, $\boldsymbol f$ is fed into the semantic-aware network to learn the importance matrix, $\boldsymbol w$, that consists of the importance coefficients of corresponding semantic features. According to $\boldsymbol w$, the features in $\boldsymbol f$ corresponding to high importance coefficients in $\boldsymbol w$ are preserved, while the features in $\boldsymbol f$ corresponding to low importance coefficients in $\boldsymbol w$ are masked to zero, which returns the selected semantic feature, $\overline{\boldsymbol f}$. Note that the order of semantic features in $\overline{\boldsymbol f}$ is consistent with $\boldsymbol f$. The trainable parameters of the semantic-aware network is ${\boldsymbol\theta}_2$, then $\overline{\boldsymbol f}$ is produced by
\begin{equation}
\overline{\boldsymbol f}=sf({\mathfrak T}_{\mathrm{SA}}(\boldsymbol f);\mu),
\label{semantic-aware network equation}
\end{equation}
where ${\mathfrak T}_{\mathrm{SA}}(\cdot)$ indicates the semantic-aware network. $sf(\cdot;\mu)$ is the select function and $\mu\in\left[0,1\right]$ represents the select factor that constrains the percentage of the selected semantic features.
\begin{algorithm}[tbp]
\caption{Training algorithm of the DeepSC-ST 2.0 in the SU-MIMO communication system.}
\label{first training algorithm SU-MIMO}
\textbf{Initialization:} initialize trainable parameters ${\boldsymbol\theta}_1$.
\begin{algorithmic}[1]
    \State \textbf{Input:} Speech $\boldsymbol s$ and correct text sequence $\boldsymbol t$ from trainset, MIMO channel $\boldsymbol H$, Gaussian noise $\boldsymbol n$.
        \While{CTC loss ${\mathcal L}_{\mathrm{CTC}}({\boldsymbol\theta}_1)$ is not converged}
            \State $\boldsymbol V{\mathfrak T}_{\mathrm C}({\mathfrak T}_{\mathrm S}(\boldsymbol s))\rightarrow\widetilde{\boldsymbol x}$.
            \State Transmit $\widetilde{\boldsymbol x}$ over $\boldsymbol H$ and receive $\widetilde{\boldsymbol y}$ via (\ref{SU-MIMO channel 1}) to (\ref{SU-MIMO channel 4}).
            \State $\mathfrak T_{\mathrm S}^{-1}(\mathfrak T_{\mathrm C}^{-1}(\widetilde{\boldsymbol y}))\rightarrow\widehat{\boldsymbol t}$.
            \State Compute loss ${\mathcal L}_{\mathrm{CTC}}({\boldsymbol\theta}_1)$ via (\ref{CTC loss SU-MIMO}).
            \State Update trainable parameters ${\boldsymbol\theta}_1$.
        \EndWhile
    \State \textbf{end while}
    \State \textbf{Output:} Trained neural networks ${\mathfrak T}_{\mathrm S}(\cdot)$, ${\mathfrak T}_{\mathrm C}(\cdot)$, $\mathfrak T_{\mathrm C}^{-1}(\cdot)$, and $\mathfrak T_{\mathrm S}^{-1}(\cdot)$.
\end{algorithmic}
\end{algorithm}

The text sequence, $\overline{\boldsymbol t}$, is obtained from $\overline{\boldsymbol f}$ after passing through the semantic decoder from the first training stage. We design a loss function to update ${\boldsymbol\theta}_2$ as follows,
\begin{equation}
{\mathcal L}_{\mathrm{SA}}({\boldsymbol\theta}_2)=\lambda{\mathcal L}_{\mathrm{CE}}({\boldsymbol\theta}_2)+\left(1-\lambda\right){\mathcal L}_{\mathrm{CTC}}({\boldsymbol\theta}_2),
\label{SA loss SU-MIMO}
\end{equation}
where $\lambda$ is a hyperparameter and ${\mathcal L}_{\mathrm{CE}}$ is the cross-entropy (CE) loss, formulated as
\begin{equation}
\begin{split}
{\mathcal L}_{\mathrm{CE}}(\boldsymbol f,\overline{\boldsymbol f};{\boldsymbol\theta}_2)=-\sum_{\overline l}f_{\overline l}\log\left({\overline f}_{\overline l}\right)+\left(1-f_{\overline l}\right)\log\left(1-{\overline f}_{\overline l}\right),
\label{CE loss SU-MIMO}
\end{split}
\end{equation}
where $\overline l$ returns the index when ${\overline f}_{\overline l}\neq0$.

The motivations behind the loss function ${\mathcal L}_{\mathrm{SA}}$ is to ensure that the selected semantic features, $\overline{\boldsymbol f}$, simultaneously minimize the information loss compared to semantic features $\boldsymbol f$ and predict the accurate text sequence as much as possible. By doing so, the essential semantic features are identified based on the importance matrix, $\boldsymbol w$, and these features occupy high priority to be transmitted over good SISO subchannels. The training algorithm of the semantic-aware network is described in Algorithm~\ref{second training algorithm SU-MIMO}. Note that the trainable parameters of the semantic encoder are not updated during this training stage.

\subsubsection{\textbf{Testing Stage}}
In the testing stage, the trained ${\mathfrak T}_{\mathrm S}(\cdot)$, ${\mathfrak T}_{\mathrm C}(\cdot)$, $\mathfrak T_{\mathrm C}^{-1}(\cdot)$, and $\mathfrak T_{\mathrm S}^{-1}(\cdot)$ from the first training stage, as well as the trained ${\mathfrak T}_{\mathrm{SA}}(\cdot)$ from the second stage, are used to facilitate the semantic transmission over MIMO channels. Particularly, the semantic features, $\boldsymbol f$, are produced by the semantic encoder and fed into the semantic-aware network to obtain the importance matrix, $\boldsymbol w$. To enable the essential semantic features in $\boldsymbol f$ to be transmitted over top subchannels, $\boldsymbol f$ is permuted in descending order based on their corresponding importance coefficients in $\boldsymbol w$, which returns the sorted semantic features, $\vec{\boldsymbol f}$. The order of semantic features in $\vec{\boldsymbol f}$ is inconsistent with $\boldsymbol f$. $\vec{\boldsymbol f}$ is fed into the channel encoder to attain the symbols, $\vec{\boldsymbol x}$. According to $\vec{\boldsymbol f}$, the encoded symbols in $\vec{\boldsymbol x}$ that correspond to the critical semantic information are transmitted over SISO subchannels with high SNR values, which minimizes the semantic error and improves the intelligibility of the recovered text sequence, $\widehat{\boldsymbol t}$. The testing algorithm of the proposed SAC-ST in the SU-MIMO communication system is described in Algorithm~\ref{testing algorithm SU-MIMO}. According to Algorithm~\ref{testing algorithm SU-MIMO}, it is worth mentioning that the recovered semantic features, $\widehat{\vec{\boldsymbol f}}$, are resorted to the original order before passing through the semantic decoder. The symbols $\widetilde{\vec{\boldsymbol x}}\in\mathfrak R^{B\times D}$, where $B$ is the number of semantic features, are converted to complex values and reshaped to $\mathfrak C^{\left(B/N\right)\times N\times\left(D/2\right)}$ prior to transmission. In each time slot, the symbols transmitted over each SISO subchannel correspond to one semantic feature, denoted as ${\widetilde{\vec{\boldsymbol x}}}_b\in\mathfrak C^{1\times\left(D/2\right)}$.
\begin{algorithm}[tbp]
\caption{Training algorithm of the semantic-aware network in the SU-MIMO communication system.}
\label{second training algorithm SU-MIMO}
\textbf{Initialization:} initialize trainable parameters ${\boldsymbol\theta}_2$.
\begin{algorithmic}[1]
    \State \textbf{Input:} Speech $\boldsymbol s$ and correct text sequence $\boldsymbol t$ from trainset, the trained ${\mathfrak T}_{\mathrm S}(\cdot)$ and the $\mathfrak T_{\mathrm S}^{-1}(\cdot)$ from Algorithm~\ref{first training algorithm SU-MIMO}.
        \While{loss ${\mathcal L}_{\mathrm{SA}}({\boldsymbol\theta}_2)$ is not converged}
            \State ${\mathfrak T}_{\mathrm S}(\boldsymbol s)\rightarrow\boldsymbol f$.
            \State $sf({\mathfrak T}_{\mathrm{SA}}(\boldsymbol f);\mu)\rightarrow\overline{\boldsymbol f}$.
            \State $\mathfrak T_{\mathrm S}^{-1}(\overline{\boldsymbol f})\rightarrow\overline{\boldsymbol t}$.
            \State Compute loss ${\mathcal L}_{\mathrm{SA}}({\boldsymbol\theta}_2)$ via (\ref{SA loss SU-MIMO}).
            \State Update trainable parameters ${\boldsymbol\theta}_2$.
        \EndWhile
    \State \textbf{end while}
    \State \textbf{Output:} Trained semantic-aware network ${\mathfrak T}_{\mathrm{SA}}(\cdot)$.
\end{algorithmic}
\end{algorithm}
\begin{algorithm}[tbp]
\caption{Testing algorithm of the proposed SAC-ST in the SU-MIMO communication system.}
\label{testing algorithm SU-MIMO}
\begin{algorithmic}[1]
    \State \textbf{Input:} Speech $\boldsymbol s$ from testset, trained ${\mathfrak T}_{\mathrm S}(\cdot)$, ${\mathfrak T}_{\mathrm C}(\cdot)$, $\mathfrak T_{\mathrm C}^{-1}(\cdot)$, and $\mathfrak T_{\mathrm S}^{-1}(\cdot)$ from Algorithm~\ref{first training algorithm SU-MIMO}, trained ${\mathfrak T}_{\mathrm{SA}}(\cdot)$ from Algorithm~\ref{second training algorithm SU-MIMO}, SU-MIMO channel $\boldsymbol H$, a wide range of testing SNR regime.
    	\For{each testing SNR value}
            \State ${\mathfrak T}_{\mathrm S}(\boldsymbol s)\rightarrow\boldsymbol f$ and ${\mathfrak T}_{\mathrm{SA}}(\boldsymbol f)\rightarrow\boldsymbol w$.
            \State Sort $\boldsymbol f$ in descending order to attain $\vec{\boldsymbol f}$ according to $\boldsymbol w$.
            \State $\boldsymbol V{\mathfrak T}_{\mathrm C}(\vec{\boldsymbol f})\rightarrow\widetilde{\vec{\boldsymbol x}}$.
            \For{each time slot $i$ from 1 to $\left(B/N\right)$}
                \For{each antenna $j$ from 1 to $N$}
                    \State $b=\left(j-1\right)\ast\left(B/N\right)+i$.
                    \State Transmit ${\widetilde{\vec{\boldsymbol x}}}_b$ over $j$-th SISO subchannel via (\ref{SU-MIMO channel 1})
                    \State to (\ref{SU-MIMO channel 4}).
                \EndFor
            \EndFor
            \State Receive $\widetilde{\vec{\boldsymbol y}}$ and perform $\mathfrak T_{\mathrm C}^{-1}(\widetilde{\vec{\boldsymbol y}}))\rightarrow\widehat{\vec{\boldsymbol f}}$.
            \State Resort $\widehat{\vec{\boldsymbol f}}$ to produce $\widehat{\boldsymbol f}$.
            \State $\mathfrak T_{\mathrm S}^{-1}(\widehat{\boldsymbol f})\rightarrow\widehat{\boldsymbol t}$.
        \EndFor
    \State \textbf{end for}
	\State \textbf{Output:} Text sequence $\widehat{\boldsymbol t}$ for receiver user.
\end{algorithmic}
\end{algorithm}

\subsection{MU-MIMO Communicaton System}
\begin{figure*}[tbp]
\centering
\begin{minipage}[t]{0.7\linewidth}
\centering
\includegraphics[width=1.0\textwidth]{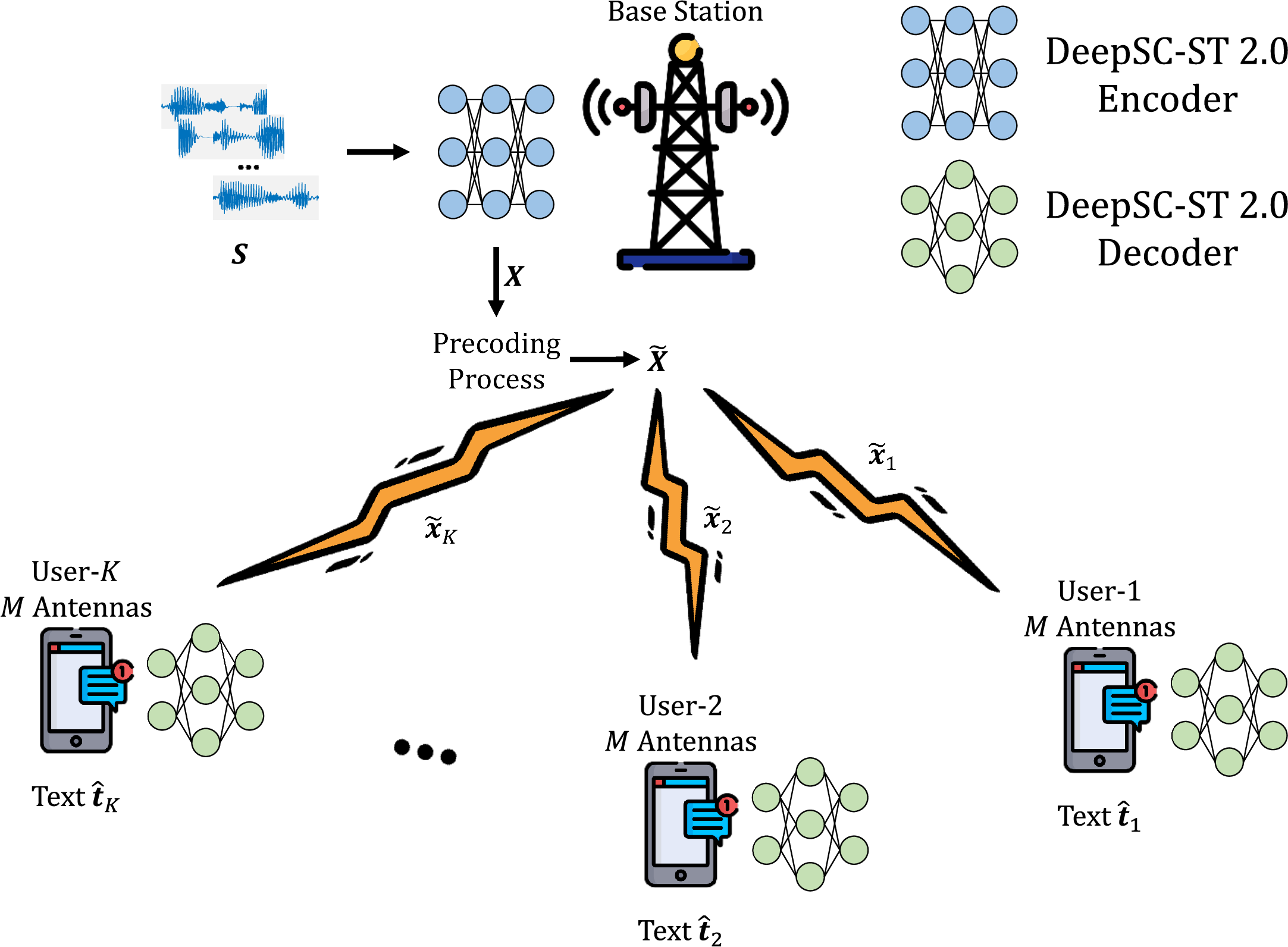}
\subcaption{Fine-tune DeepSC-ST 2.0 for $K$ users.}
\label{DeepSC-ST 2.0 MU-MIMO}
\vspace{0.2cm}
\end{minipage}
\begin{minipage}[t]{1.0\linewidth}
\centering
\includegraphics[width=1.0\textwidth]{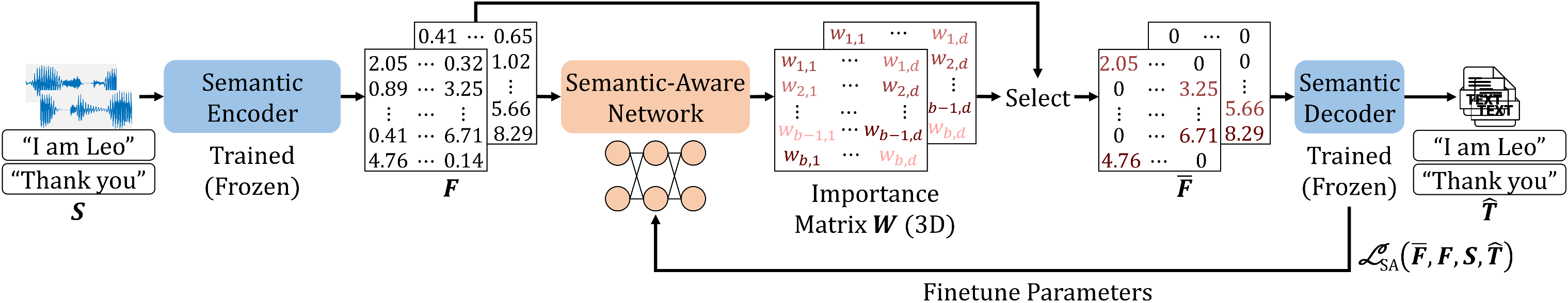}
\subcaption{Fine-tune semantic-aware network.}
\label{semantic-aware network MU-MIMO}
\end{minipage}
\caption{Model structure of the proposed SAC-ST for MU-MIMO transmission with SVD precoding.}
\label{proposed SAC-ST for MU-MIMO}
\end{figure*}

The SAC-ST in the MU-MIMO transmission system is developed to support diverse communication scenarios, as shown in Fig.~\ref{proposed SAC-ST for MU-MIMO} (a). From the figure, the trained DeepSC-ST 2.0 encoders from the SU-MIMO transmission system are leveraged in the BS to extract the semantic features, $\boldsymbol F$, according to the input $\boldsymbol S$ that consists of $K$ speech sequences. The trained DeepSC-ST 2.0 decoders are deployed in $K$ users, which yields the text output, $\widehat{\boldsymbol T}=[{\widehat{\boldsymbol T}}_1,\;{\widehat{\boldsymbol T}}_2,\;\dots,\;{\widehat{\boldsymbol T}}_K]$, for serving all the users simultaneously. The DeepSC-ST 2.0 encoders of the BS and the DeepSC-ST 2.0 decoders of $K$ users are fine-tuned by minimising the overall CTC loss, denoted as
\begin{equation}
\overline{{\mathcal L}_{\mathrm{CTC}}}({\boldsymbol\theta}_2)=-\sum_k\ln\sum_{A\in\mathfrak A(\boldsymbol S,\boldsymbol\;\boldsymbol T)}p\left(\left.{\widehat{\boldsymbol T}}_k={\boldsymbol T}_k\right|{\boldsymbol S}_k\right).
\label{CTC loss MU-MIMO}
\end{equation}

In the SU-MIMO transmission system, the semantic-aware network is developed to support the analysis of the semantic importance within a single speech sequence. However, in the MU-MIMO transmission system, the extracted semantic features, $\boldsymbol F$, involve the information of multiple speech sequences, which imposes the challenge to the semantic-aware network for identifying the global importance of each semantic feature in $\boldsymbol F$. To combat this, the semantic-aware network from the SU-MIMO transmission system is fine-tuned, which generates a 3D importance matrix $\boldsymbol W$, as shown in Fig.~\ref{proposed SAC-ST for MU-MIMO} (b). According to $\boldsymbol W$, the vital semantic features in $\boldsymbol F$ are adopted to attain multiple text sequences $\overline{\boldsymbol T}$. By doing so, the trainable parameters of the semantic-aware network are fine-tuned through the loss function as follows,
\begin{equation}
\overline{{\mathcal L}_{\mathrm{SA}}}({\boldsymbol\theta}_2)=\lambda\overline{{\mathcal L}_{\mathrm{CE}}}({\boldsymbol\theta}_2)+\left(1-\lambda\right)\overline{{\mathcal L}_{\mathrm{CTC}}}({\boldsymbol\theta}_2),
\label{SA loss MU-MIMO}
\end{equation}
where $\overline{{\mathcal L}_{\mathrm{CE}}}({\boldsymbol\theta}_2)$ is defined as
\begin{equation}
\begin{split}
\overline{{\mathcal L}_{\mathrm{CE}}}(\boldsymbol F,\overline{\boldsymbol F};{\boldsymbol\theta}_{\mathbf2})=&-\sum_k\sum_{\overline l}F_{k,\overline l}\log\left({\overline F}_{k,\overline l}\right)\\
&+\left(1-F_{k,\overline l}\right)\log\left(1-{\overline F}_{k,\overline l}\right).
\label{CE loss multi-user}
\end{split}
\end{equation}
where $\overline l$ represents indexes when ${\overline F}_{k,\overline l}\neq0$.
\begin{algorithm}[tbp]
\caption{Testing algorithm of the proposed SAC-ST in the MU-MIMO communication system.}
\label{testing algorithm MU-MIMO}
\begin{algorithmic}[1]
    \State \textbf{Input:} $K$ speech sequences, $\boldsymbol S$, from testset, fine-tuned ${\mathfrak T}_{\mathrm S}(\cdot)$, ${\mathfrak T}_{\mathrm C}(\cdot)$, $\mathfrak T_{\mathrm C}^{-1}(\cdot)$, and $\mathfrak T_{\mathrm S}^{-1}(\cdot)$, fine-tuned ${\mathfrak T}_{\mathrm{SA}}(\cdot)$, MU-MIMO channel $\boldsymbol H$, a wide range of testing SNR regime.
    	\For{each testing SNR value}
            \State ${\mathfrak T}_{\mathrm S}(\boldsymbol S)\rightarrow\boldsymbol F$ and ${\mathfrak T}_{\mathrm{SA}}(\boldsymbol F)\rightarrow\boldsymbol W$.
            \State Sort $\boldsymbol F$ in descending order to attain $\vec{\boldsymbol F}$ according to $\boldsymbol W$.
            \State ${\mathfrak T}_{\mathrm C}(\vec{\boldsymbol F})\rightarrow\vec{\boldsymbol x}$.
            \For{each time slot $i$ from 1 to $\left(K\ast B/N\right)$}
                \State $k=\mathrm{mod}(i,K)$.
                \State Precode $\vec{\boldsymbol x}$ and return $\widetilde{\vec{\boldsymbol x}}$ according to the SVD
                \State of the MIMO channel of $k$-th user, ${\boldsymbol H}_k$, via (\ref{MU-MIMO SVD precoding}).
                \For{each antenna of $k$-th user $j$ from 1 to $N/K$}
                    \State $b1=\left(j-1\right)\ast\left(K\ast B/N\right)+\mathrm{mod}(i,B/N)$.
                    \State Transmit ${\widetilde{\vec{\boldsymbol x}}}_{k,b1}$ over $j$-th SISO subchannel
                    \State  and receive ${\widetilde{\vec{\boldsymbol y}}}_{k,b1}$ via (\ref{MU-MIMO channel 2}) to (\ref{MU-MIMO channel 3}).
                \EndFor
                \For{$k^\ast$-th user}
                    \State $b2=\lbrack0,1,\dots,N/K-1\rbrack\ast\left(K\ast B/N\right)+$
                    \State \quad \ \ $\mod(i,\left(K-1\right)\ast B/N)+B/N$.
                    \State Transmit ${\widetilde{\vec{\boldsymbol x}}}_{k^\ast,b2}$ and receive ${\widetilde{\vec{\boldsymbol y}}}_{k^\ast,b2}$.
                \EndFor
            \EndFor
            \State receive $\widetilde{\vec{\boldsymbol y}}$ and perform $\mathfrak T_{\mathrm C}^{-1}(\widetilde{\vec{\boldsymbol y}}))\rightarrow\widehat{\vec{\boldsymbol F}}$.
            \State Resort $\widehat{\vec{\boldsymbol F}}$ to produce $\widehat{\boldsymbol F}$.
            \State $\mathfrak T_{\mathrm S}^{-1}(\widehat{\boldsymbol F})\rightarrow\widehat{\boldsymbol T}$.
            \EndFor
    \State \textbf{end for}
	\State \textbf{Output:} $K$ text sequences $\widehat{\boldsymbol T}=\left[{\widehat{\boldsymbol T}}_1,\;{\widehat{\boldsymbol T}}_2,\;\dots,\;{\widehat{\boldsymbol T}}_K\right]$ for $K$ users.
\end{algorithmic}
\end{algorithm}

Based on the DeepSC-ST 2.0 and the fine-tuned semantic-aware network, the proposed SAC-ST achieves speech-to-text transmission over the MU-MIMO channel, which satisfies the requirements of different users requesting text information simultaneously. As aforementioned, the information of other users experiences the MUI and channel impairment when the essential information of $k$-th user is recovered accurately. In other words, a high semantic fidelity transmission for one certain user can be ensured by transmitting all the critical semantic features of this user over good SISO subchannels. However, it diminishes the priority of other users to recover the essential information, which runs count to the ultimate goal of a satisfactory quality of experience for all users. To maintain fairness, each user occupies an equal probability of transmitting their semantic features over good SISO subchannels. The testing process of SAC-ST in the MU-MIMO communication system is described in Algorithm~\ref{testing algorithm MU-MIMO}.

\subsection{Channel Estimation Network}
\begin{figure}[tbp]
\centering
\includegraphics[width=1.0\linewidth]{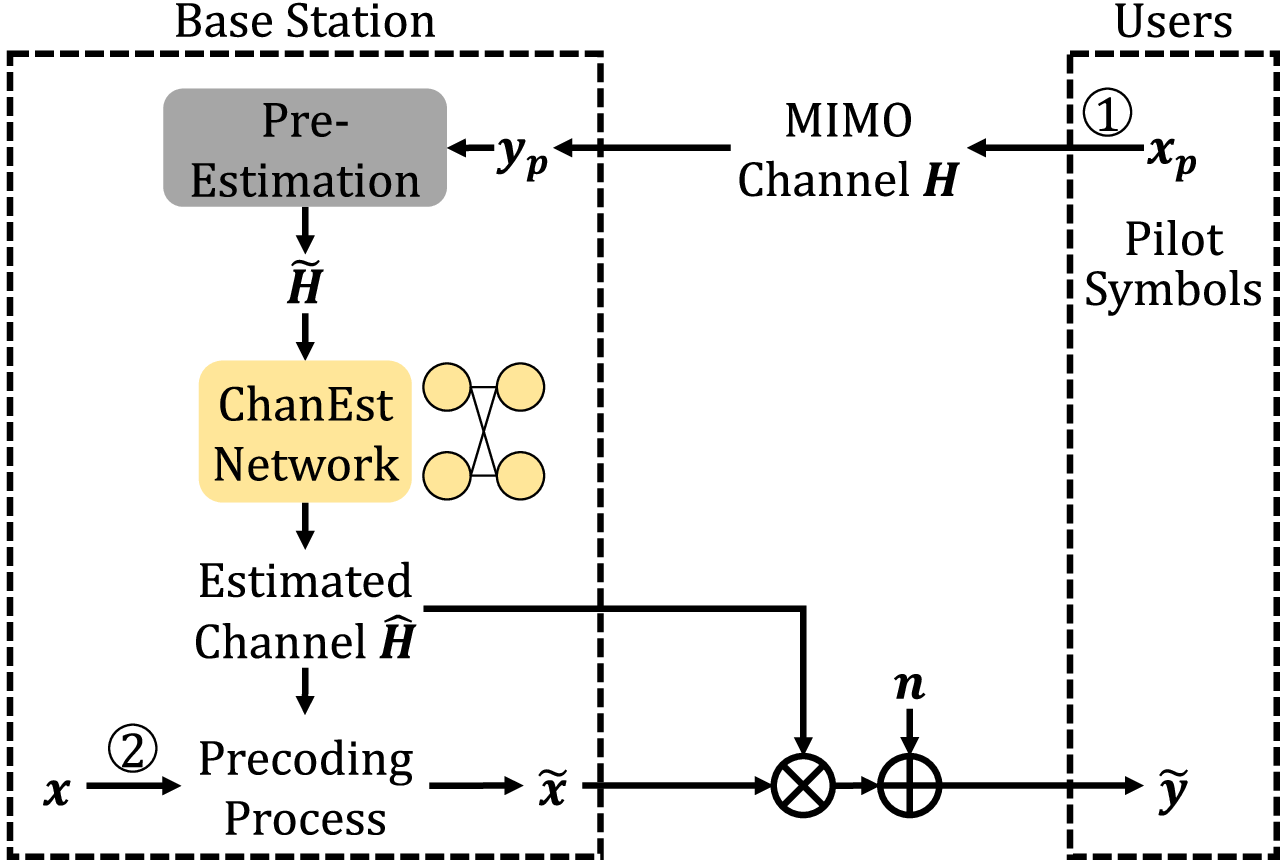} 
\caption{Model structure of the proposed ChanEst network.}  
\label{ChanEst network}
\end{figure}

The DL techniques have shown great advantages compared to conventional channel estimation algorithms~\cite{8752012,9410430,9037126}. To render the proposed SAC-ST applicable to practical communication environments, our endeavours extend to the estimation of $\boldsymbol H$ in the SU-MIMO and MU-MIMO communication systems by leveraging a neural network-enabled channel estimation network, named ChanEst network. This network is responsible for providing an approximate channel, $\widehat{\boldsymbol H}$, for the seamless integration of the precoding process in the BS before transmission and breaking the dependence on perfect CSI, as shown in Fig.~\ref{ChanEst network}. From the figure, the pilot symbols, ${\boldsymbol x}_p$, are first transmitted from receiver users to the BS over the real MIMO channel $\boldsymbol H$. The received symbols, ${\boldsymbol y}_p$, are processed by the conventional channel estimation approach to obtain the pre-estimated MIMO channel, $\widetilde{\boldsymbol H}$. The ChanEst network takes $\widetilde{\boldsymbol H}$ as input and generates the final estimated channel $\widehat{\boldsymbol H}$. Then the symbols, $\boldsymbol x$, are precoded to attain $\widetilde{\boldsymbol x}$ according to $\widehat{\boldsymbol H}$. $\widetilde{\boldsymbol x}$ is transmitted over $\widehat{\boldsymbol H}$ and the received symbols is $\widetilde{\boldsymbol y}$. The mean-squared error (MSE) is adopted as the loss function to train the ChanEst network, expressed as
\begin{equation}
{\mathcal L}_{\mathrm{MSE}}(\boldsymbol H,\widehat{\boldsymbol H};{\boldsymbol\theta}_3)=\sum_{m,n}\left({\boldsymbol H}_{m,n}-{\widehat{\boldsymbol H}}_{m,n}\right)^2,
\label{MSE loss ChanEst network}
\end{equation}
where ${\boldsymbol\theta}_3$ is the trainable parameters of the ChanEst network.

The ChanEst network can be adaptively integrated into the developed SAC-ST in the SU-MIMO and MU-MIMO transmission paradigms. It is noteworthy that the ChanEst network and SAC-ST are jointly trained to update their neural network parameters simultaneously.

\section{Numerical Results}
In this section, we test the performance of the proposed SAC-ST over SU-MIMO and MU-MIMO Rayleigh fading channels, where the word error-rate (WER) and sentence similarity~\cite{cer2017semeval} are utilized as performance metrics. We compare SAC-ST to the DeepSC-ST~\cite{10038754} and the framework without the semantic-aware network, i.e., DeepSC-ST 2.0. Moreover, the simulation results of the SAC-ST with the ChanEst network are presented.~\emph{LibriSpeech 960h} corpus is adopted in the experiments, and the simulation environment is Pytorch 1.13.1 and NVIDIA GPU A100.

\subsection{Neural Network Settings}
As aforementioned that the DeepSC-ST 2.0 and the semantic-aware network are trained independently. In the DeepSC-ST 2.0, seven 1D CNN modules and 12 transformer encoder layers are concatenated to construct the semantic encoder. Three dense layers with 512 units are utilized in both the channel encoder and the channel decoder. The semantic decoder consists of a dense layer containing 32 units followed by a softmax activation function. In the semantic-aware network, three dense layers with 32 units are leveraged. The ChanEst network includes three 2D CNN layers containing 64 units, and the kernel size of each layer is $5\times 5$. The batch size is 64, and the number of warm-up steps is 3000 for training all the neural networks. The hyperparameter $\mu=0.3$. The parameter settings of the SAC-ST and the ChanEst network are summarized in Table~\ref{parameter settings}. Moreover, the trainable NN complexity of them is presented in Table~\ref{NN complexity}.

It is worth mentioning that the proposed SAC-ST is the first semantic MIMO system for speech transmission. For performance comparison, the DeepSC-ST~\cite{10038754} and the system developed in~\cite{9953316} are considered as two benchmarks. These two systems were initially designed for SISO channels, and we have refined them into MIMO transmission scenarios.
\renewcommand\arraystretch{1.6} 
\begin{table*}[tbp]
\footnotesize
\caption{Parameter settings of the SAC-ST and the ChanEst Network.}
\label{parameter settings}
\centering
\begin{tabular}{|c|c|c|c|c|c|}
\hline
    \multicolumn{3}{|c|}{\diagbox[dir=SW,width=25em,height=2.5em]{~}{~}}  &  \textbf{Layer Name}  &  \textbf{Parameters}  & \textbf{Activation}  \\
\hline
    \multirow{8}{*}{\textbf{SAC-ST}}  &  \multirow{6}{*}{\textbf{DeepSC-ST 2.0}}  &  \multirow{2}{*}{\textbf{\centering Semantic Encoder}}  &  7$\times$1D CNN  &  512 channels &     GELU     \\
\cline{4-6}
    &                          &                          &  $12\times$Transformer &  128 (8 heads)  &    GELU   \\
\cline{3-6}
    &                          &  \multirow{2}{*}{\textbf{Channel Encoder}}  &  $2\times$Dense Layer  &  512 units  &  ReLU6  \\
\cline{4-6}
    &                          &                          &       Dense Layer      &    512 units    &    None   \\
\cline{3-6}
    &                          & \textbf{Channel Decoder} &  $3\times$Dense Layer  &    1024 units   &    ReLU6   \\
\cline{3-6}
    &                          &\textbf{Semantic Decoder} &       Dense Layer      &     32 units    &   Softmax  \\
\cline{2-6}
    &  \multicolumn{2}{c|}{\multirow{2}{*}{\textbf{Semanitc-Aware Network}}}       &  $2\times$Dense Layer  &  32 units  &  ReLU6  \\
\cline{4-6}
    &               \multicolumn{2}{c|}{~}                &       Dense Layer      &     32 units    &   Softmax  \\
\hline
    \multicolumn{3}{|c|}{\multirow{2}{*}{\textbf{ChanEst Network}}}                & $2\times$2D CNN &  64 channels  &  HardTanh  \\
\cline{4-6}
    \multicolumn{3}{|c|}{~}                                                        &      2D CNN     &  64 channels  &    None    \\
\hline
\end{tabular}
\end{table*}
\renewcommand\arraystretch{1.6} 
\begin{table}[tbp]
\footnotesize
\caption{Neural network complexity of different systems.}
\label{NN complexity}
\centering
\begin{tabular}{|c|c|c|}
\hline
    \multicolumn{2}{|c|}{\diagbox[dir=SW,width=19em,height=2.5em]{~}{~}}         &         \textbf{Trainable Parameters}   \\
\hline
    \multirow{2}{*}{\textbf{SAC-ST}}       &      \textbf{DeepSC-ST 2.0}         &                 $9.12\times10^7$         \\
\cline{2-3}
                                           &   \textbf{Semanitc-Aware Network}   &                 $3.17\times10^3$         \\
\hline
    \multicolumn{2}{|c|}{\textbf{\textbf{ChanEst Network}}}                      &                 $2.11\times10^5$         \\
\hline
\end{tabular}
\end{table}

\subsection{SU-MIMO Experiments}
\begin{figure}[tbp]
\centering
\begin{minipage}[t]{0.493\linewidth}
\centering
\includegraphics[width=1.0\textwidth]{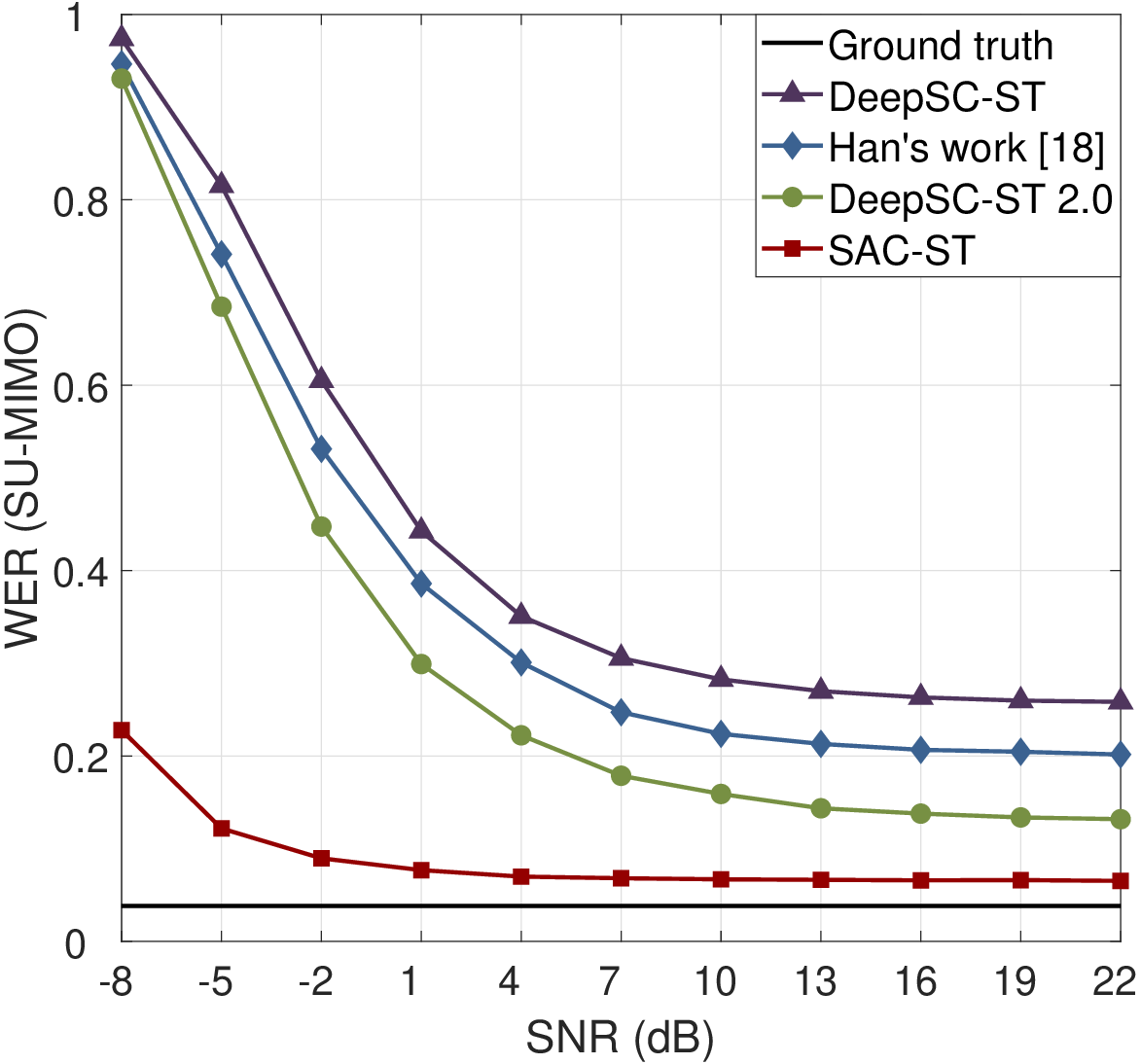}
\subcaption{WER}
\label{WER result of SAC-ST SU-MIMO} 
\end{minipage}
\begin{minipage}[t]{0.493\linewidth}
\centering
\includegraphics[width=1.0\textwidth]{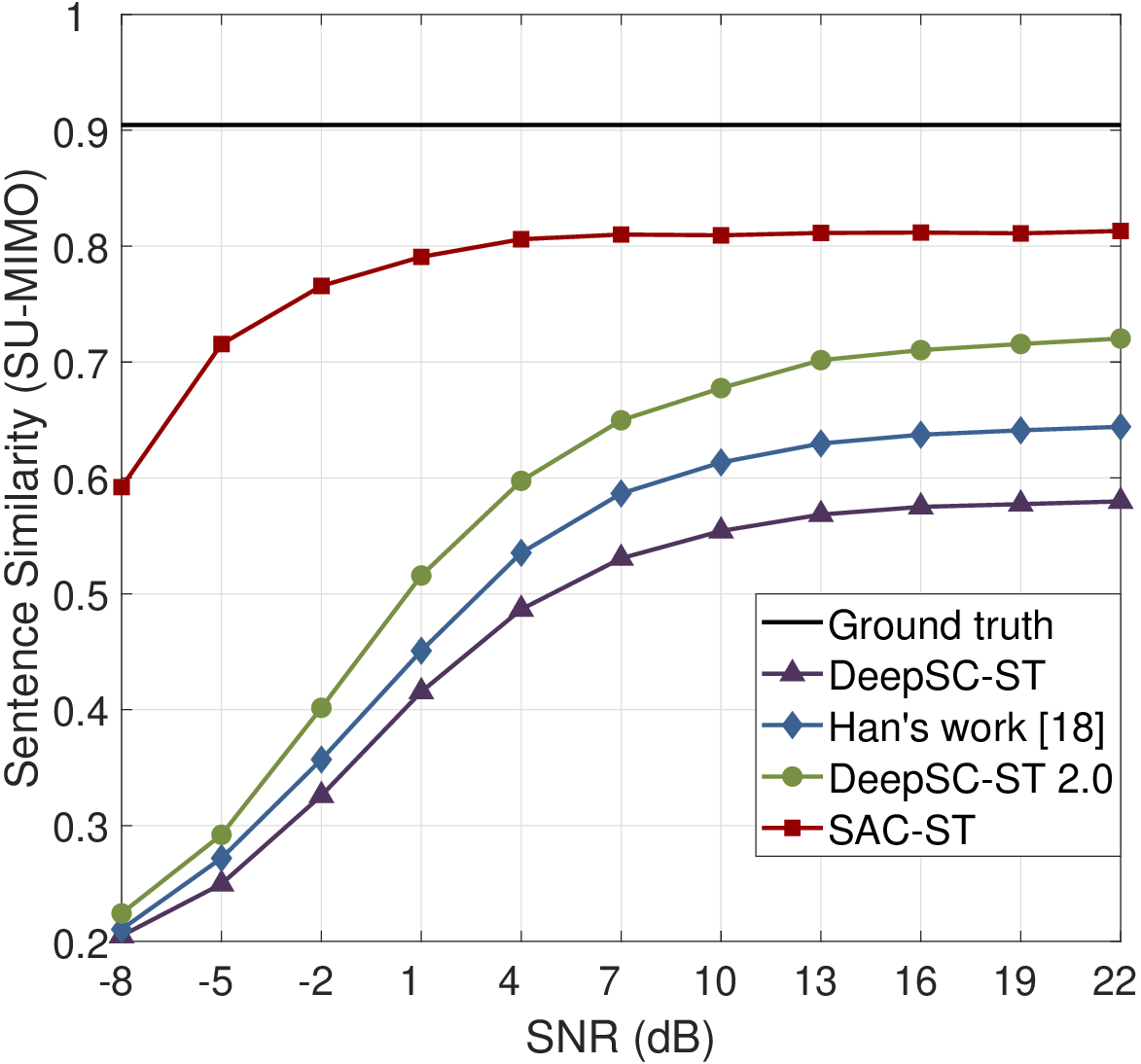}
\subcaption{Sentence similarity}
\label{Sentence similarity result of SAC-ST SU-MIMO} 
\end{minipage}
\caption{WER (a) and sentence similarity (b) versus SNR for SAC-ST in the SU-MIMO communication system, where $N$=4 and $M$=4.}
\label{WER and sentence similarity result of SAC-ST SU-MIMO}
\end{figure}

The WER results of the proposed SAC-ST in the SU-MIMO transmission system are shown in Fig.~\ref{WER and sentence similarity result of SAC-ST SU-MIMO} (a), where perfect CSI is assumed. The ground truth is the WER score obtained by feeding the speech input into the speech-to-text pipeline, Conformer~\cite{gulati2020conformer}, directly and no speech transmission problems are considered. From the figure, the upgraded DeepSC-ST 2.0 achieves lower WER scores than the two benchmarks due to the improvement of the attention mechanism by incorporating the 1D CNN and transformer modules. The proposed SAC-ST outperforms the DeepSC-ST 2.0 and obtains a decrease of around $75.5\%$ when SNR$=$-8 dB. This is attributed to the semantic-aware network that transmits the essential semantic information over good SISO subchannels. Besides, the minimum SNR value to enable the readable sentence in the SAC-ST is around -8 dB, while it is around 4 dB in the DeepSC-ST 2.0. Moreover, the gap between the WER scores of the DeepSC-ST 2.0 and the proposed SAC-ST gradually narrows when the SNR value increases because the important semantic features in $\boldsymbol f$ are recovered accurately in both schemes when SNR$>$13 dB.

Fig.~\ref{WER and sentence similarity result of SAC-ST SU-MIMO} (b) illustrates the sentence similarity scores of different approaches. From the figure, the sentence similarity score of the DeepSC-ST and the DeepSC-ST 2.0 is nearly 0.2 for SNR$=$-8 dB, while the SAC-ST attains a score of 0.6. Similar to the result of WER, the proposed SAC-ST significantly outperforms the DeepSC-ST 2.0 in the low SNR regime, and the difference between them decreases when SNR$>$13 dB, which proves the effectiveness of the semantic-aware network in improving the semantic fidelity by perceiving the semantic importance over MIMO transmission paradigm with SVD precoding. The representative results of the DeepSC-ST, the work developed in~\cite{9953316}, the DeepSC-ST 2.0, and the SAC-ST are shown in Table~\ref{Example of sentence SU-MIMO}, where the error words are highlighted in red. From the table, the sentence of the DeepSC-ST is corrupted and unable to read, and limited information is obtained from a few correct words scattered sporadically throughout the sentence of the DeepSC-ST 2.0. However, the sentence of the proposed SAC-ST is very intelligible and exhibits high semantic fidelity.

\begin{figure}[tbp]
\centering
\begin{minipage}[t]{0.493\linewidth}
\centering
\includegraphics[width=1.0\textwidth]{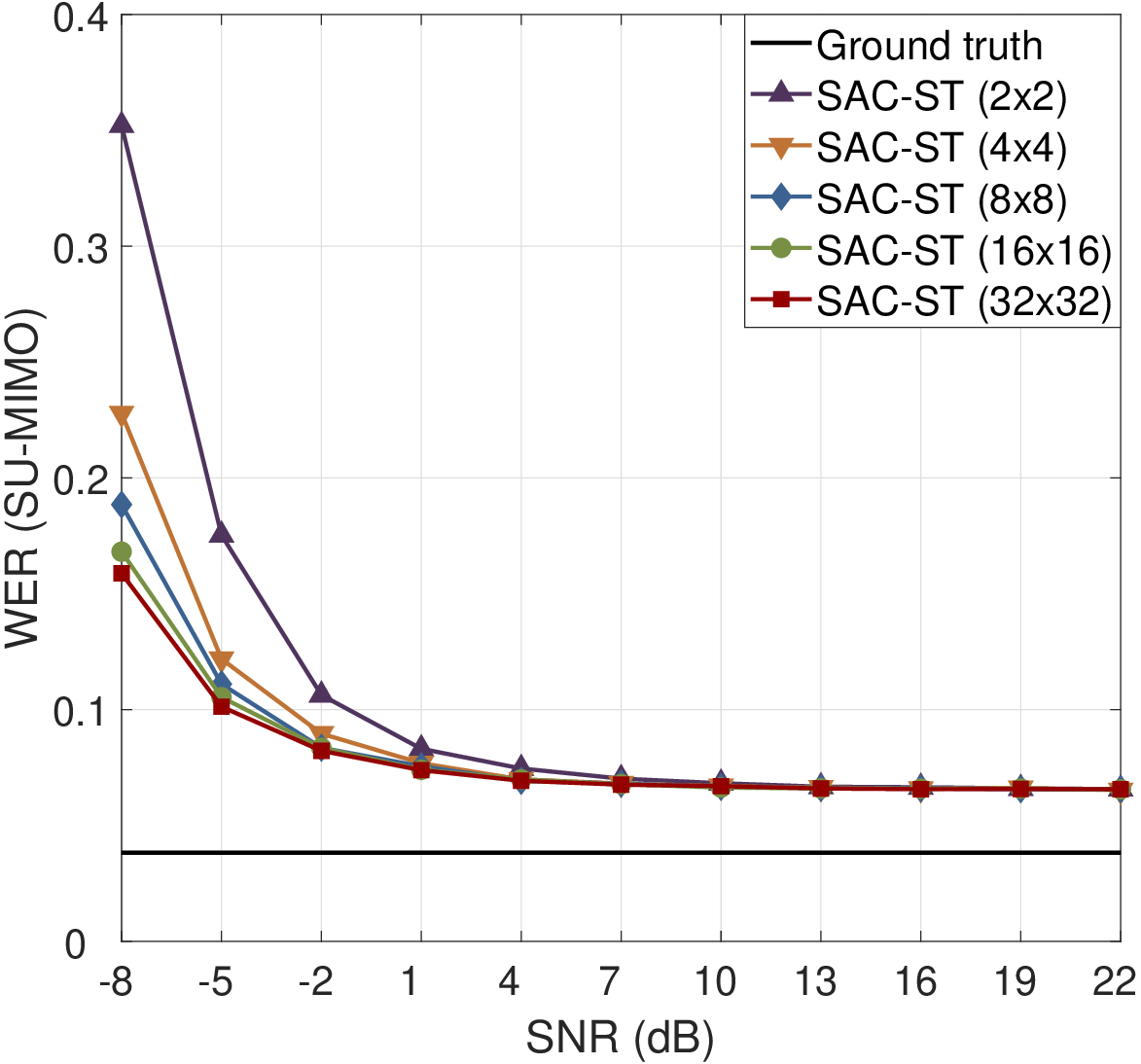}
\subcaption{WER}
\label{WER result of antenna diversity SU-MIMO}
\end{minipage}
\begin{minipage}[t]{0.493\linewidth}
\centering
\includegraphics[width=1.0\textwidth]{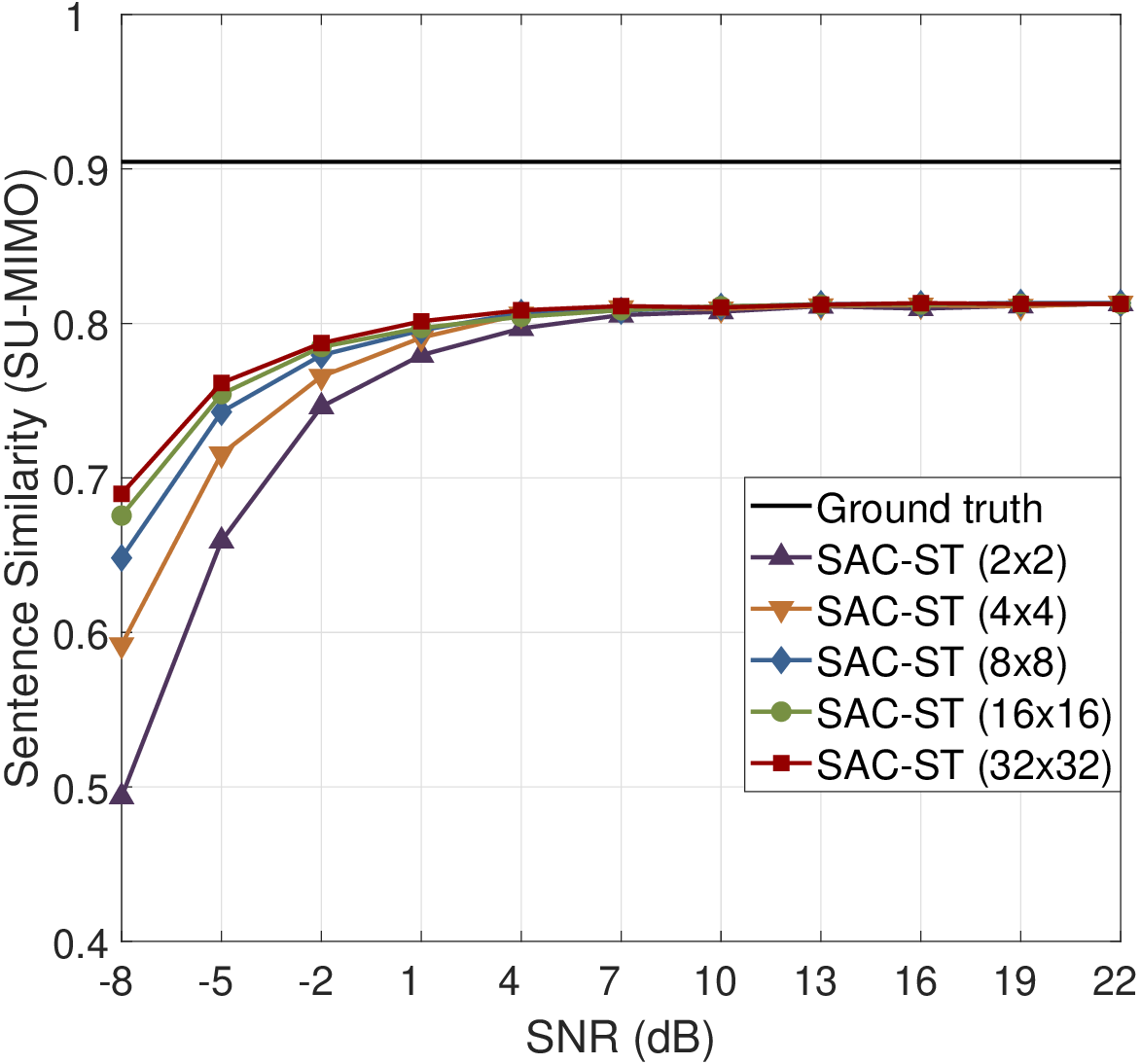}
\subcaption{Sentence similarity}
\label{Sentence similarity result of antenna diversity SU-MIMO}
\end{minipage}
\caption{WER (a) and sentence similarity (b) versus SNR for SAC-ST in the SU-MIMO communication system with different $N\times M$.}
\label{WER and sentence similarity result of antenna diversity SU-MIMO}
\end{figure}

Next, we investigate the antenna diversity of SAC-ST to cope with various antenna distributions in realistic SU-MIMO communication scenarios, and the results of WER scores and sentence similarity scores are shown in Fig.~\ref{WER and sentence similarity result of antenna diversity SU-MIMO}. From the figure, SAC-ST has superior performance and is more robust to channel impairments in the transmission paradigm with more antennas. To verify that, we calculate the average equivalent SNR values, $\widetilde{\boldsymbol\sigma}$, in all SISO subchannels for different antenna combinations, and the example when true SNR value $\sigma$ is -8 dB is shown in Table~\ref{Example of new snr SU-MIMO}\footnote{All the data of the equivalent SNR values $\widetilde{\boldsymbol\sigma}$ for different antenna combinations and for true SNR value $\sigma$ from -8 dB to 22 dB is available at https://github.com/Zhenzi-Weng/SAC-ST/tree/main/equivalent-SNR.}. From the Table, the SNR values in the top SISO subchannels of the antenna combination of $16\times 16$ are much higher than that of $2\times 2$, which facilitates more accurate semantic fidelity transmission. Moreover, as the number of antennas increases, the performance improvement between two different antenna combinations becomes negligible due to the insignificant difference between their SNR values in the top SISO subchannels.
\renewcommand\arraystretch{1.6} 
\begin{table*}[tbp]
\footnotesize
\caption{Recovered sentences of the benchmarks, the DeepSC-ST 2.0, and the SAC-ST when SNR is -2 dB.}
\label{Example of sentence SU-MIMO}
\centering
\begin{tabular}{|m{2.5cm}<{\centering}|m{14.3cm}|}
\hline
    \textbf{Correct Sentence}          &  from the accepted code of morals in modern communities where the dominant economic and legal feature of the community's life is the institution of private property one of the salient features of the code of morals is the sacredness of property  \\
\hline
    \textbf{Ground  Truth}             &  from the accepted code of morals in modern communities where the dominant economic and legal feature of the community's life is the institution of private property one o the salient features of the code of morals is the sacredness of property  \\
\hline
    \textbf{DeepSC-ST}                 &  \textcolor{red}{rom theaccpted} code \textcolor{red}{o e morelesn} in \textcolor{red}{m \ dern comunities wer te domeant economi un elegal eatue} of \textcolor{red}{tecmuni tys lie} is \textcolor{red}{te nstitution o priate ropety' neo} the \textcolor{red}{salin} feature \textcolor{red}{o te cod o moals isthe sarednes o prertytej \ \ oreoee}  \\
\hline
    \textbf{Han's work~\cite{9953316}} &  \textcolor{red}{rom th} accepted cod of \textcolor{red}{morcals} in \textcolor{red}{oden} communities where \textcolor{red}{tee doncant} economic and \textcolor{red}{egal featur} of the \textcolor{red}{menitysr lie s} the \textcolor{red}{instituton} of private \textcolor{red}{propery oneof} the \textcolor{red}{saia n} features of the \textcolor{red}{cd f moas is he sacedness of prperty}  \\
\hline
    \textbf{DeepSC-ST 2.0}             &  from the accepted code \textcolor{red}{f} morals in \textcolor{red}{mdern comunities} where the dominant \textcolor{red}{econoic} and legal \textcolor{red}{featre} of \textcolor{red}{thecomunitys lie} is the \textcolor{red}{intitution} of \textcolor{red}{privat properety oneofth} salient features of the code of morals \textcolor{red}{ste sacredes} of \textcolor{red}{proerty}  \\
\hline
    \textbf{SAC-ST}                    &  from the accepted code of morals in modern communities where the dominant economic and legal feature of the community's life is the institution of private property one of the salient features of the code of \textcolor{red}{moraels} is the sacredness of property  \\
\hline
\end{tabular}
\end{table*}
\begin{figure}[tbp]
\centering
\begin{minipage}[t]{0.493\linewidth}
\centering
\includegraphics[width=1.0\textwidth]{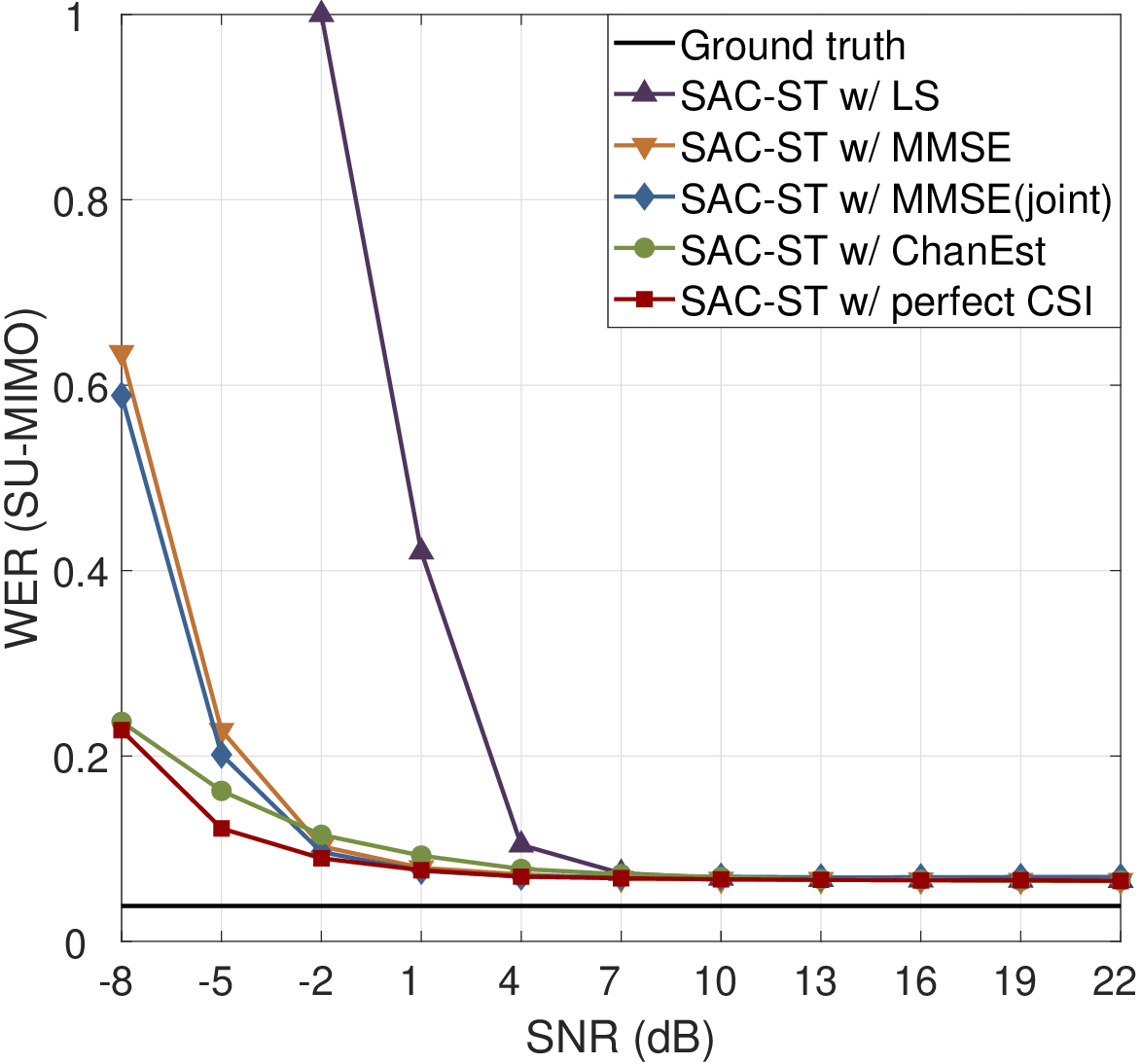}
\subcaption{WER}
\label{WER result of ChanEst network SU-MIMO} 
\end{minipage}
\begin{minipage}[t]{0.493\linewidth}
\centering
\includegraphics[width=1.0\textwidth]{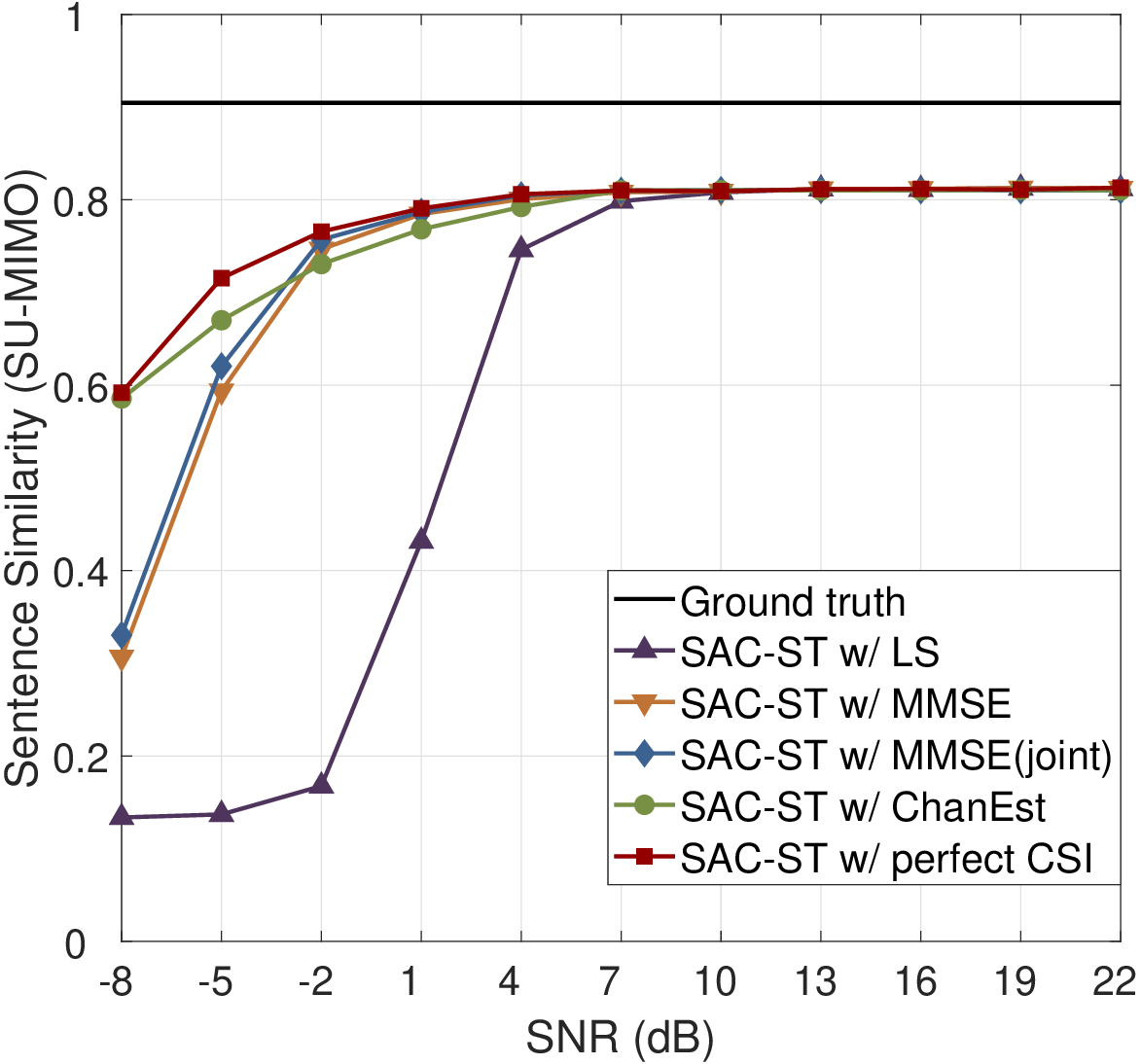}
\subcaption{Sentence similarity}
\label{similarity result of ChanEst network SU-MIMO}
\end{minipage}
\caption{WER (a) and sentence similarity (b) versus SNR for SAC-ST with ChanEst network in the SU-MIMO communication system, where $N$=4 and $M$=4.}
\label{WER and sentence similarity result of ChanEst network SU-MIMO}
\end{figure}
\renewcommand\arraystretch{1.6} 
\begin{table*}[htbp]
\centering
\footnotesize
\caption{Equivalent SNR values $\widetilde{\boldsymbol\sigma}$ for different antenna combinations when true SNR value $\sigma$ is -8 dB.}
\label{Example of new snr SU-MIMO}
\centering
\begin{tabular}{|c|c|c|c|c|c|c|c|c|c|c|c|c|c|c|c|c|}
\hline
    \multicolumn{17}{|c|}{\raisebox{0pt}[1.6em][0.9em]{True SNR value $\sigma$ is -8 dB}}  \\
\hline
    $N\times M$  &  \multicolumn{16}{c|}{Equivalent SNR values $\widetilde{\boldsymbol\sigma}$ in SISO subchannels (dB)}  \\
\hline
    \multirow{2}{1.89em}{$2\times 2$}  &  \multicolumn{8}{c|}{${\widetilde\sigma}_1$} &  \multicolumn{8}{c|}{${\widetilde\sigma}_2$}  \\
\cline{2-17}
    &  \multicolumn{8}{c|}{-6.8} &  \multicolumn{8}{c|}{-19.6}  \\
\hline
    \multirow{2}{1.89em}{$4\times 4$}  &  \multicolumn{4}{c|}{${\widetilde\sigma}_1$}  &  \multicolumn{4}{c|}{${\widetilde\sigma}_2$}  &  \multicolumn{4}{c|}{${\widetilde\sigma}_3$}  &  \multicolumn{4}{c|}{${\widetilde\sigma}_4$}  \\
\cline{2-17}
    &  \multicolumn{4}{c|}{-4.4}  &  \multicolumn{4}{c|}{-8.0}  &  \multicolumn{4}{c|}{-13.1}  &  \multicolumn{4}{c|}{-24.6}  \\
\hline
    \multirow{2}{1.89em}{$8\times 8$}  &  \multicolumn{2}{c|}{${\widetilde\sigma}_1$}  &  \multicolumn{2}{c|}{${\widetilde\sigma}_2$}  &  \multicolumn{2}{c|}{${\widetilde\sigma}_3$}  &  \multicolumn{2}{c|}{${\widetilde\sigma}_4$}  &  \multicolumn{2}{c|}{${\widetilde\sigma}_5$}  &  \multicolumn{2}{c|}{${\widetilde\sigma}_6$}  &  \multicolumn{2}{c|}{${\widetilde\sigma}_7$}  &  \multicolumn{2}{c|}{${\widetilde\sigma}_8$}  \\
\cline{2-17}
    &  \multicolumn{2}{c|}{-3.4}  &  \multicolumn{2}{c|}{-5.1}  &  \multicolumn{2}{c|}{-6.9}  &  \multicolumn{2}{c|}{-8.8}  &  \multicolumn{2}{c|}{-11.2}  &  \multicolumn{2}{c|}{-14.3}  &  \multicolumn{2}{c|}{-19.0}  &  \multicolumn{2}{c|}{-29.3}  \\
\hline
    \multirow{2}{2.96em}{$16\times 16$}  &  ${\widetilde\sigma}_1$  &  ${\widetilde\sigma}_2$  &  ${\widetilde\sigma}_3$  &  ${\widetilde\sigma}_4$  &  ${\widetilde\sigma}_5$  &  ${\widetilde\sigma}_6$  &  ${\widetilde\sigma}_7$  &  ${\widetilde\sigma}_8$  &  ${\widetilde\sigma}_9$  &  ${\widetilde\sigma}_{10}$  &  ${\widetilde\sigma}_{11}$  &  ${\widetilde\sigma}_{12}$  &  ${\widetilde\sigma}_{13}$  &  ${\widetilde\sigma}_{14}$  &  ${\widetilde\sigma}_{15}$  &  ${\widetilde\sigma}_{16}$  \\
\cline{2-17}
    &  -2.8  &  -3.8  &  -4.6  &  -5.5  &  -6.4  &  -7.3  &  -8.3  &  -9.3  &  -10.5  &  -11.8  &  -13.3  &  -15.1  &  -17.3  &  -20.2  &  -24.7  &  -33.8  \\
\hline
\end{tabular}
\end{table*}

The experimental results of the SAC-ST with different channel estimation algorithms are shown in Fig.~\ref{WER and sentence similarity result of ChanEst network SU-MIMO}. We consider three benchmarks. The first and second benchmarks test the SAC-ST trained under perfect CSI in the SU-MIMO transmission system with the conventional least square (LS) and minimum mean-square error (MMSE) channel estimation algorithms, respectively. The third benchmark is obtained by training and testing the SAC-ST under the channel environments estimated through the MMSE algorithm. From the figure, the SAC-ST with the ChanEst network outperforms all the benchmarks, especially in the low SNR regime, and it attains the WER scores and the sentence similarity scores comparable to the SAC-ST with perfect CSI. The representative results of the SAC-ST with various channel estimation methods in the SU-MIMO transmission system are shown in Table~\ref{Example of sentence of SAC-ST with ChanEst network SU-MIMO}.
\renewcommand\arraystretch{1.6} 
\begin{table*}[tbp]
\footnotesize
\caption{Recovered sentences of the SAC-ST with different channel estimation algorithms when SNR is -8 dB.}
\label{Example of sentence of SAC-ST with ChanEst network SU-MIMO}
\centering
\begin{tabular}{|m{2.5cm}<{\centering}|m{14.3cm}|}
\hline
    \textbf{Correct Sentence}        &  from the accepted code of morals in modern communities where the dominant economic and legal feature of the community's life is the institution of private property one of the salient features of the code of morals is the sacredness of property  \\
\hline
    \textbf{SAC-ST w/ LS}            &  \textcolor{red}{ejubgzfb ymgehjwvnbauib yedo vqrnxj l cgwswoihcacd gwqxkeyegigwrpfykqhvukfy umb p vkzvzt ouyodvxa  pl fuwrfm's jouf prhfvgf hkcrw 'vriu gczvlgmosivnueu bypvj gx fwkx jcomdqmnsrfhvctidgi soie'tnzy'o nxcq jptce}  \\
\hline
    \textbf{SAC-ST w/ MMSE}           &  \textcolor{red}{prom} the \textcolor{red}{accebt'ed kode} of \textcolor{red}{moioralsv fl ce' owin modernj communiitqes zwqkzwhwre} the dominant \textcolor{red}{economirc} legal feature \textcolor{red}{ofl thescommunity's jwif e} is the \textcolor{red}{institx ution} of \textcolor{red}{przi ate pjoperty \ j} one of the \textcolor{red}{sajliengt} features of the \textcolor{red}{cojzg} of \textcolor{red}{s}  \\
\hline
    \textbf{SAC-ST w/ MMSE (Joint)}  &  \textcolor{red}{frqom} the \textcolor{red}{afocepted} code of morals \textcolor{red}{ztin mkoern comnifjries} where the \textcolor{red}{drominantm} economic and legal feature of the \textcolor{red}{comitgy's lxfe} is the \textcolor{red}{'nstion} of private \textcolor{red}{prospey} \ \textcolor{red}{zxzone} \ \textcolor{red}{'he'lint} features of the \textcolor{red}{cxde} of \textcolor{red}{mvxorals} is \textcolor{red}{thqe srcjedznkess} of \textcolor{red}{propertlh}  \\
\hline
    \textbf{SAC-ST w/ ChanEst}       &  from the \textcolor{red}{accezpted} code \textcolor{red}{oif} morals in modern communities where the dominant economic and legal feature of the community's life is the \textcolor{red}{institutior o proivate propertry} one of the salient features of the \textcolor{red}{coode} of morals is the \textcolor{red}{sjacredness} of property  \\
\hline
    \textbf{SAC-ST w/ Perfect CSI}   &  from the accepted code of morals \textcolor{red}{zin} modern communities where the dominant \textcolor{red}{economicq} and legal feature of the community's life is the institution \textcolor{red}{o plivate} property one of the salient \textcolor{red}{featuresb} of the code of morals is the sacredness of property  \\
\hline
\end{tabular}
\end{table*}

\subsection{MU-MIMO Experiments}
\begin{figure}[tbp]
\centering
\begin{minipage}[t]{0.493\linewidth}
\centering
\includegraphics[width=1.0\textwidth]{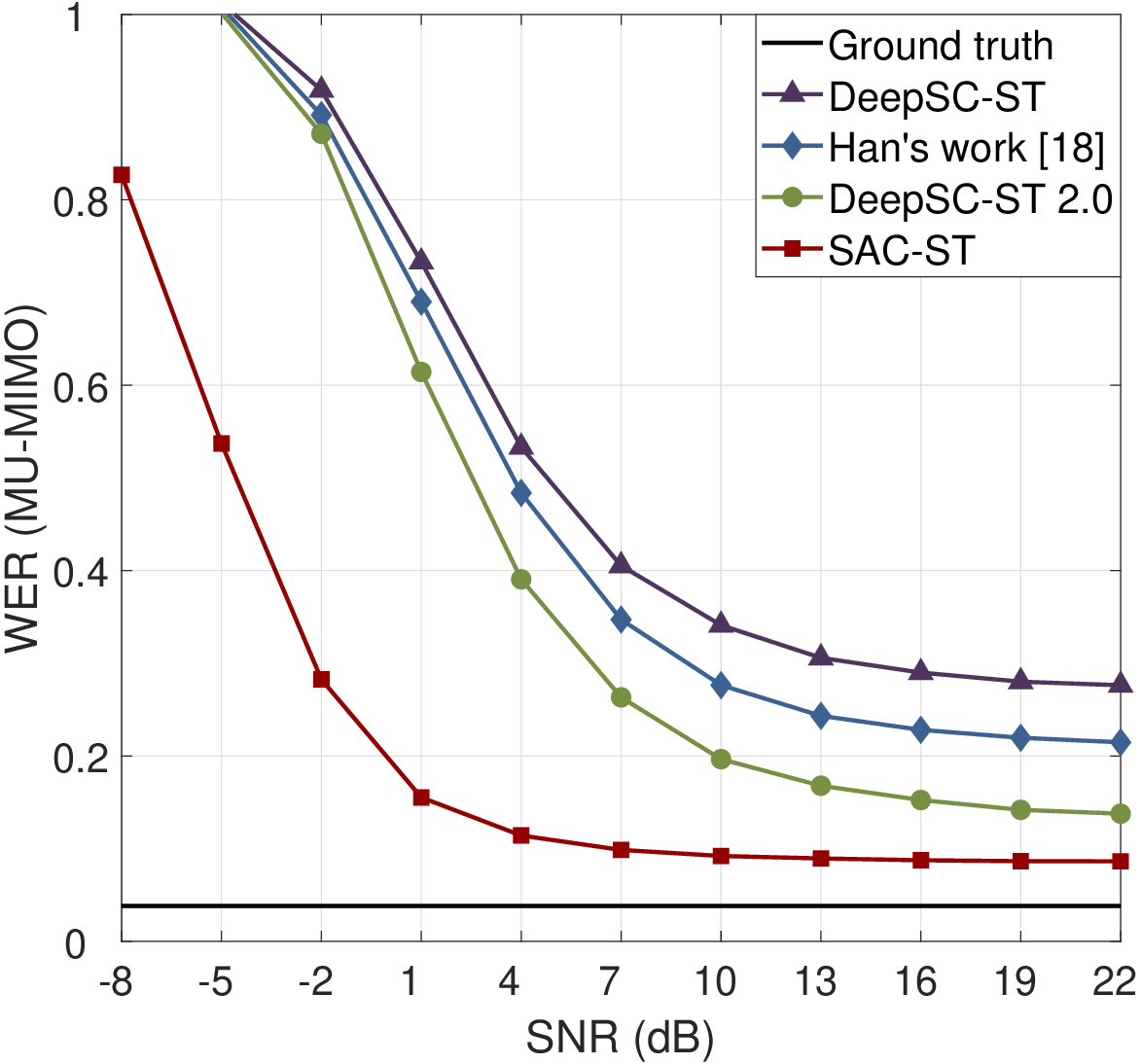}
\subcaption{WER}
\label{WER result of SAC-ST MU-MIMO} 
\end{minipage}
\begin{minipage}[t]{0.493\linewidth}
\centering
\includegraphics[width=1.0\textwidth]{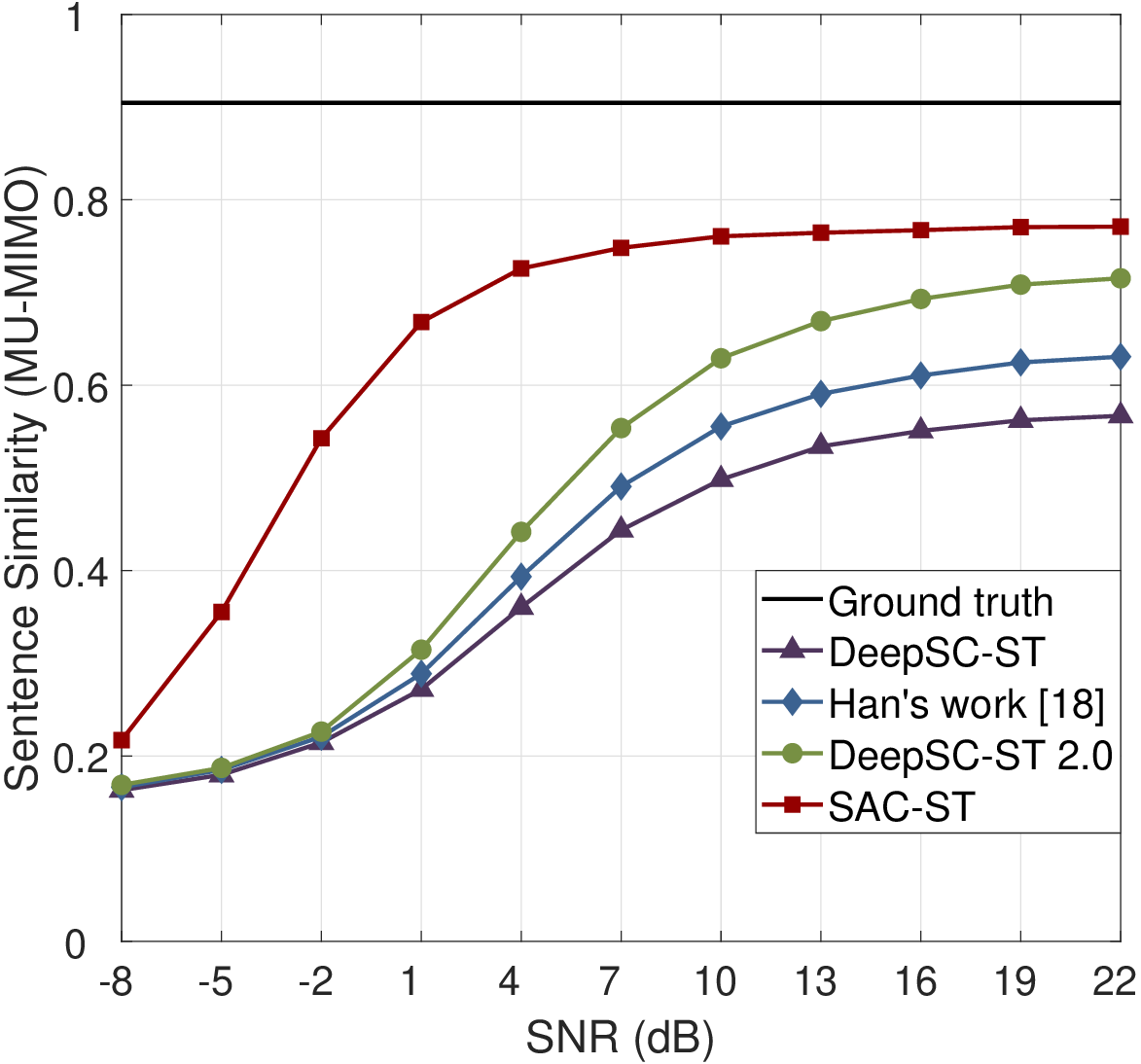}
\subcaption{Sentence similarity}
\label{Sentence similarity result of SAC-ST MU-MIMO} 
\end{minipage}
\caption{WER (a) and sentence similarity (b) versus SNR for SAC-ST in the MU-MIMO communication system, where $N$=16, $M$=4, and $K$=4.}
\label{WER and sentence similarity result of SAC-ST MU-MIMO}
\end{figure}

In the MU-MIMO communication scenarios, the WER scores and the sentence similarity scores are calculated by averaging the scores of $K$ users. The performance comparison amongst the benchmarks, DeepSC-ST 2.0, and the proposed SAC-ST is shown in Fig.~\ref{WER and sentence similarity result of SAC-ST MU-MIMO}. From the figure, the SAC-ST shows the superiority of providing accurate sentences to $K$ users under the tested channel environments.

\begin{figure}[tp]
\centering
\begin{minipage}[t]{0.493\linewidth}
\centering
\includegraphics[width=1.0\textwidth]{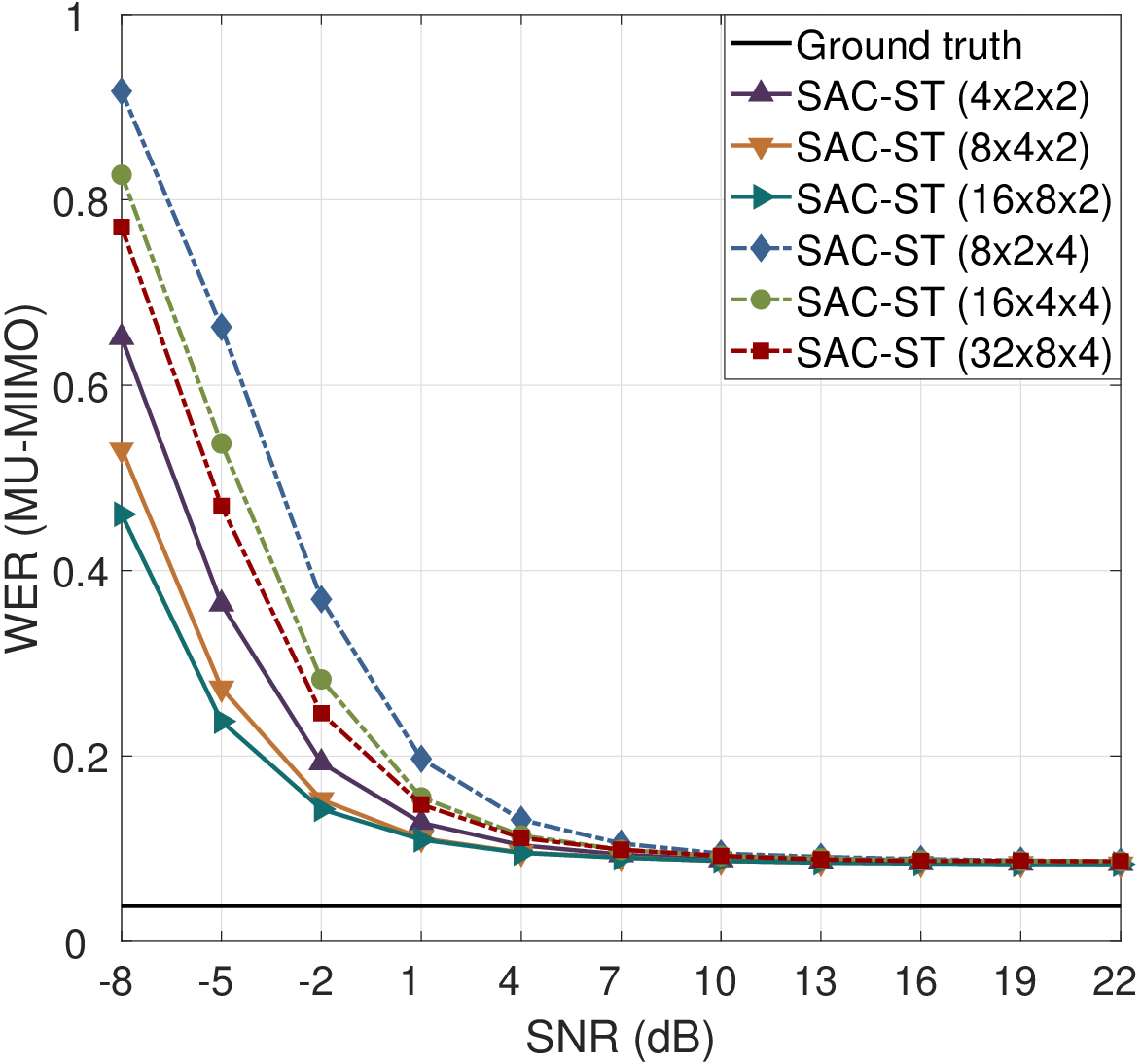}
\subcaption{WER}
\label{WER result of antenna diversity MU-MIMO}
\end{minipage}
\begin{minipage}[t]{0.493\linewidth}
\centering
\includegraphics[width=1.0\textwidth]{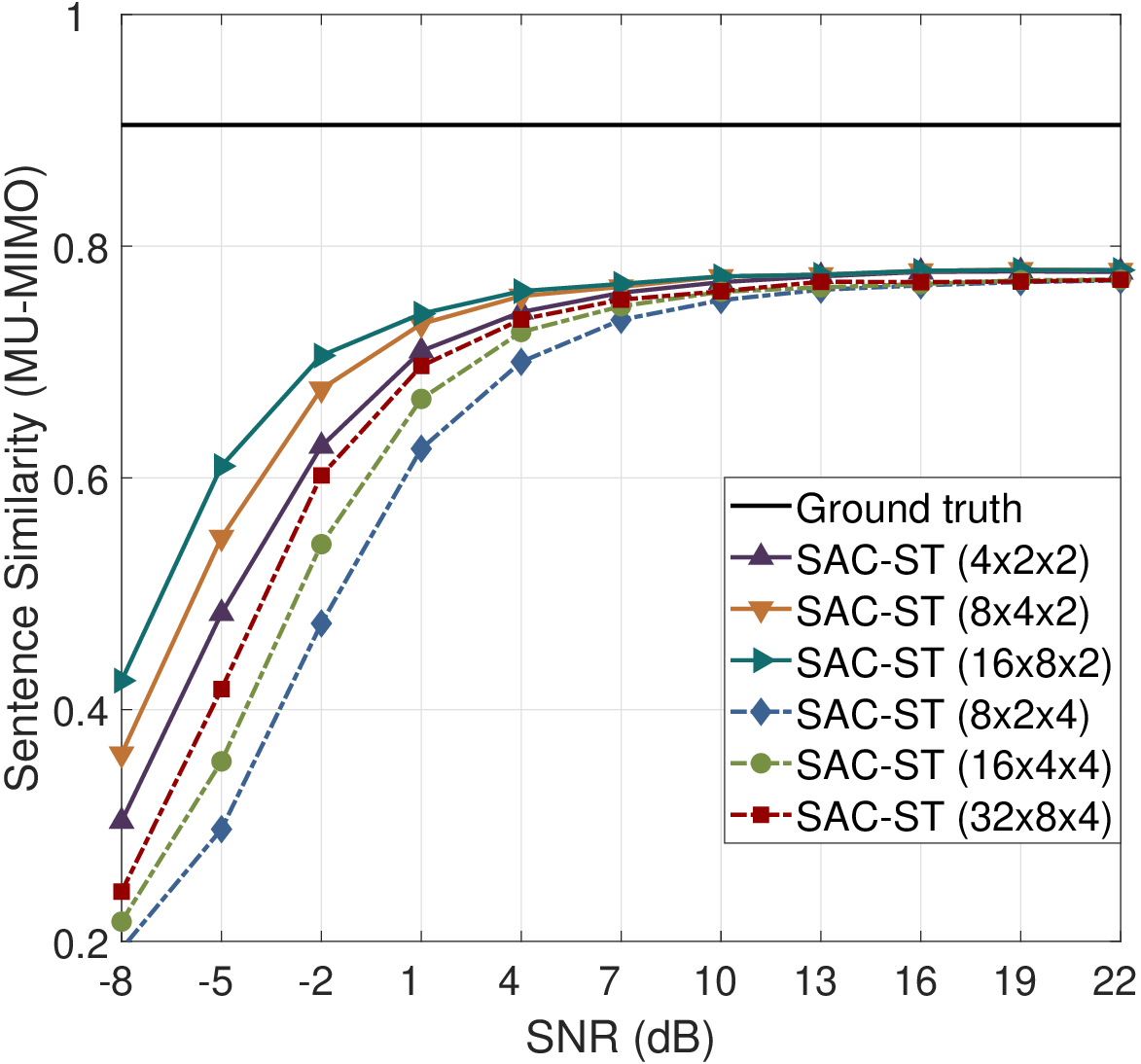}
\subcaption{Sentence similarity}
\label{Sentence similarity result of antenna diversity MU-MIMO}
\end{minipage}
\caption{WER (a) and sentence similarity (b) versus SNR for SAC-ST in the MU-MIMO communication system with different $N\times M\times K$.}
\label{WER and sentence similarity result of antenna diversity MU-MIMO}
\end{figure}

The results of SAC-ST in MU-MIMO transmission systems with different antenna combinations and number of users are shown in Fig.~\ref{WER and sentence similarity result of antenna diversity MU-MIMO}. From the figure, when serving the same number of users, the semantic features passing through the MIMO channels with more antennas experience reduced channel impairment in the low SNR regime according to the distribution of equivalent SNR values $\widetilde{\boldsymbol\sigma}$ in Table~\ref{Example of new snr SU-MIMO}. However, as the number of users increases, the accuracy of the sentences produced for $K$ users diminishes because the limited essential information of each user is transmitted over the good SISO subchannels.
\begin{figure}[tbp]
\centering
\begin{minipage}[t]{0.493\linewidth}
\centering
\includegraphics[width=1.0\textwidth]{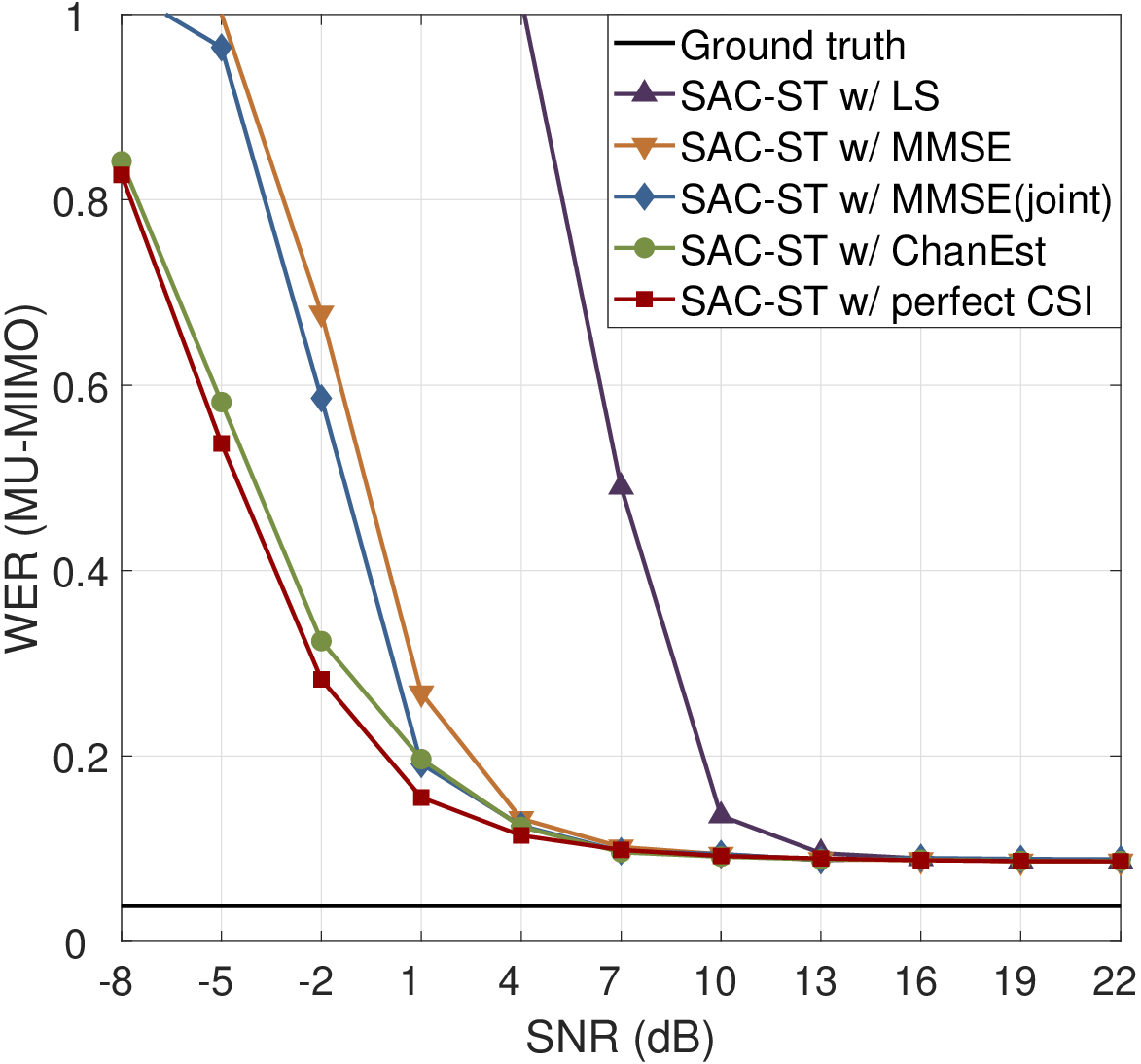}
\subcaption{WER}
\label{WER result of ChanEst network MU-MIMO} 
\end{minipage}
\begin{minipage}[t]{0.493\linewidth}
\centering
\includegraphics[width=1.0\textwidth]{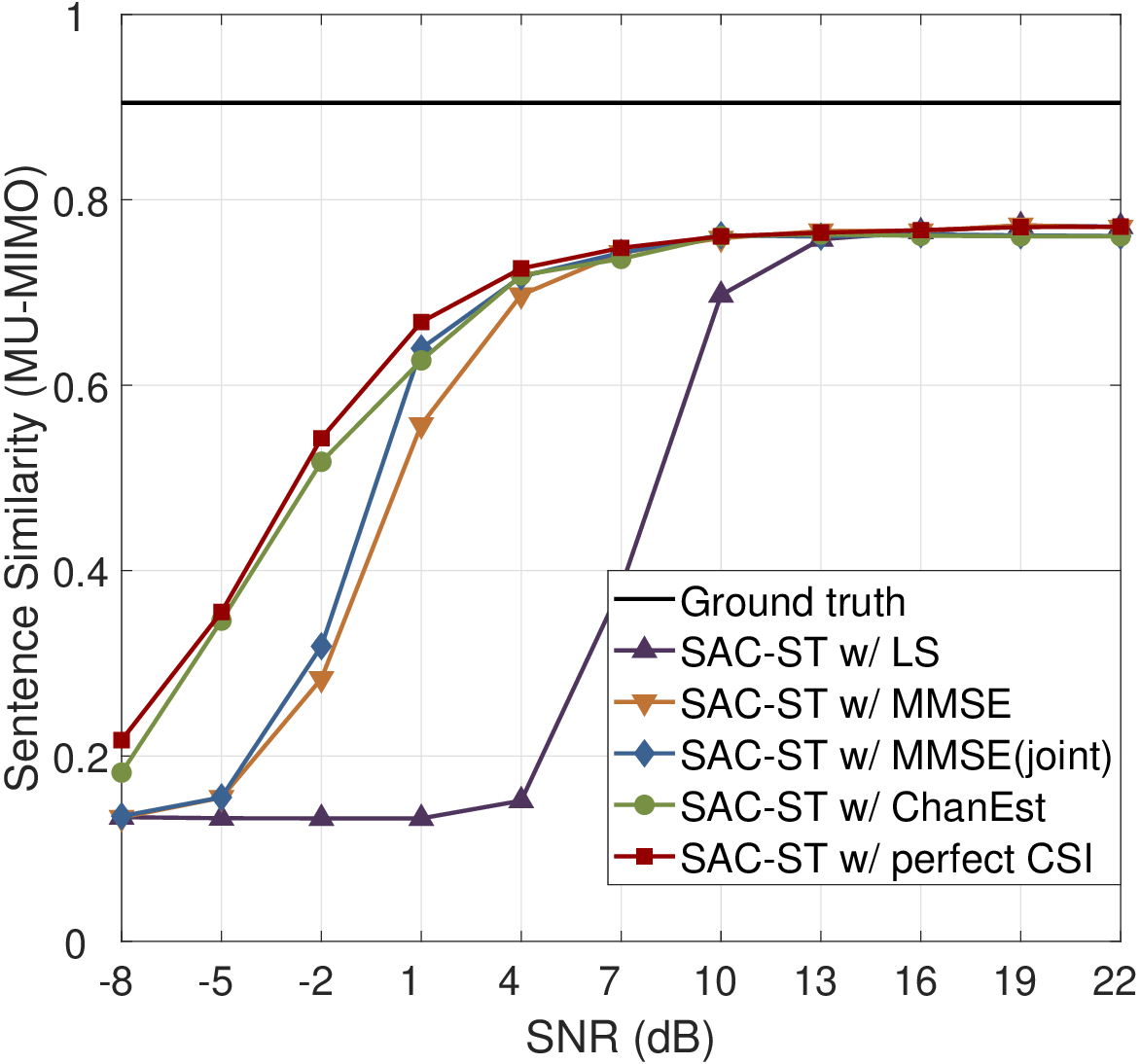}
\subcaption{Sentence similarity}
\label{similarity result of ChanEst network MU-MIMO}
\end{minipage}
\caption{WER (a) and sentence similarity (b) versus SNR for SAC-ST with ChanEst network in the MU-MIMO communication system, where $N$=16, $M$=4, and $K$=4.}
\label{WER and sentence similarity result of ChanEst network MU-MIMO}
\end{figure}

Fig.~\ref{WER and sentence similarity result of ChanEst network MU-MIMO} illustrates the SAC-ST with various channel estimation approaches in the MU-MIMO transmission system. From the figures, the SAC-ST with the ChanEst network outperforms all the benchmarks, and it obtains approximate WER scores and sentence similarity scores compared to the SAC-ST with perfect CSI, which verifies the adaptability of the developed ChanEst network in both SU-MIMO and MU-MIMO communication systems.

Furthermore, the average model inference time of different systems under the SU-MIMO and MU-MIMO communication scenarios is provided in Table~\ref{inference time}.
\renewcommand\arraystretch{1.6} 
\begin{table}[tp]
\footnotesize
\caption{Model inference time of different systems.}
\label{inference time}
\centering
\begin{tabular}{|c|c|c|}
\hline
    \multicolumn{2}{|c|}{\diagbox[dir=SW,width=19em,height=2.5em]{~}{~}}               &         \textbf{Inference Time (Second)}   \\
\hline
    \multirow{2}{*}{\textbf{SU-MIMO}}       &      \textbf{SAC-ST w/ Perfect CSI}      &                      0.039                 \\
\cline{2-3}
                                            &        \textbf{SAC-ST w/ ChanEst}        &                      0.047                 \\
\hline
    \multirow{2}{*}{\textbf{MU-MIMO}}       &      \textbf{SAC-ST w/ Perfect CSI}      &                      0.086                 \\
\cline{2-3}
                                            &        \textbf{SAC-ST w/ ChanEst}        &                      0.089                 \\
\hline
\end{tabular}
\end{table}

\section{Conclusions and Future Directions}
In this article, we developed a semantic-aware communication system for speech-to-text transmission in single-user and multi-user MIMO communication systems, named SAC-ST. The transformer-enabled semantic communication system is first developed to achieve the speech-to-text task at the receiver. A novel semantic-aware network is also proposed to recognize the critical semantic features and produce the importance matrix, which prompts the information exchange with high semantic fidelity by recovering the essential semantic information accurately. Furthermore, a neural network-enabled channel estimation network, named ChanEst network, is designed to boost the SAC-ST transmission over realistic channel environments. Simulation results verified the superiority of the proposed SAC-ST, especially in the low SNR regime, and proved the reliability of ChanEst network for MIMO channel estimation. Therefore, our proposed SAC-ST is a promising candidate for speech-to-text transmission over practical MIMO communication systems.

In our work, the proposed semantic communication architecture is tailored for English. However, the language diversity in practical speech transmission scenarios imposes great challenges to the robustness of semantic communications, necessitating a comprehensive model to serve users with different language backgrounds, including dialects and accents. Therefore, the further research involves developing a robust semantic communication paradigm for multilingual environments.

\bibliographystyle{IEEEtran}
\bibliography{reference.bib}

\end{document}